\documentclass[10pt,journal]{IEEEtran}
\usepackage{amsmath}
\usepackage{amsfonts}
\usepackage{epsfig}
\usepackage{multirow}
\usepackage{xcolor}
\usepackage{rotating}
\usepackage[normalem]{ulem}

\usepackage[utf8]{inputenc}
\usepackage{booktabs}
\usepackage{mdframed}
\usepackage{url}
\usepackage{mathtools}

\usepackage[all]{xy}
\usepackage{tabu}

\graphicspath{{./figures/}}

 \def\0z{{\boldsymbol{0}}}
 \def\1z{{\boldsymbol{1}}}
 \def\2z{{\boldsymbol{2}}}
 \def\3z{{\boldsymbol{3}}}
 \def\4z{{\boldsymbol{4}}}
 \def\5z{{\boldsymbol{5}}}
 \def\6z{{\boldsymbol{6}}}
 \def\7z{\boldsymbol{7}}
 \def\8z{\boldsymbol{8}}
 \def\9z{\boldsymbol{9}}

 \def\uz{{\boldsymbol{u}}}
 
 \def\wz{{\boldsymbol{w}}}
 \def\xz{{\boldsymbol{x}}}
 \def\yz{{\boldsymbol{y}}}

 \def\muz{{\boldsymbol{\mu}}}

 \def\Lambdaz{{\boldsymbol{\Lambda}}}

 \def\xx{{\text{x}}}

\newcommand{\UU}{{\mathbf U}}

\newcommand{\WW}{{\mathbf W}}

\newcommand{\CC}{{\mathbf C}}
\newcommand{\AAA}{{\mathbf A}}
\newcommand{\BBB}{{\mathbf B}}
\newcommand{\XX}{{\mathbf X}}
\newcommand{\YY}{{\mathbf Y}}

\newcommand{\II}{{\mathbf I}}

\newcommand{\cblue}{\textcolor{black}}

\DeclareMathOperator{\Trc}{Tr}

\begin{document}

\title{Non-Negative OPLS for Supervised Design of Filter Banks: Application to Image and Audio Feature Extraction}

\author{Sergio~Mu\~noz-Romero,
        Jer\'onimo Arenas-Garc\'ia,~\IEEEmembership{Senior Member,~IEEE},
        and~Vanessa G\'omez-Verdejo
\thanks{S. Mu\~noz-Romero was with the Department of Signal Theory and Communications, Universidad Rey Juan Carlos, 28943, Fuenlabrada, Spain, and with the Center for Computational Simulation, Universidad Polit\'ecnica de Madrid, Spain; e-mail: sergio.munoz@urjc.es. \mbox{ph. +34 91 624 8759}.}
\thanks{J. Arenas-Garc\'ia and V. G\'omez-Verdejo are with the Department of Signal Theory and Communications, Universidad Carlos III de Madrid, 28911 Legan\'es, Spain; e-mails: \{jarenas,vanessa\}@tsc.uc3m.es.}
\thanks{This work has been partly supported by MINECO projects TEC2013-48439-C4-1-R, TEC2014-52289-R, TEC2016-75161-C2-1-R, TEC2016-75161-C2-2-R, TEC2016-81900-REDT/AEI, and PRICAM (S2013/ICE-2933).}}



\IEEEcompsoctitleabstractindextext{
\begin{abstract}
Audio or visual data analysis tasks usually have to deal with high-dimensional and non-negative signals. However, most data analysis methods suffer from overfitting and numerical problems when data have more than a few dimensions needing a dimensionality reduction preprocessing. 
Moreover, \cblue{interpretability about how and why filters work for audio or visual applications is a desired property, specially when energy or spectral signals are involved. In these cases,  due to the nature of these signals, the non-negativity of the filter weights is a desired property to better understand its working.}
Because of these two necessities, we propose different methods to reduce the dimensionality of data while the non-negativity and interpretability of the solution are assured. 
In particular, we propose a generalized methodology to design filter banks in a supervised way \cblue{for applications dealing with} non-negative data, and we explore different ways of solving the proposed \cblue{objective} function consisting of a non-negative version of \cblue{Orthonormalized Partial Least Squares (OPLS)} method. We analyze the discriminative power of the features obtained with the proposed methods for two different and widely studied applications: texture and music genre classification. Furthermore, we compare the filter banks achieved by our methods with other state-of-the-art methods  \cblue{specifically designed} for feature extraction.
\end{abstract}

}

\maketitle
\IEEEdisplaynotcompsoctitleabstractindextext
\IEEEpeerreviewmaketitle

\ifCLASSOPTIONcompsoc
  \noindent\raisebox{2\baselineskip}[0pt][0pt]%
  {\parbox{\columnwidth}{\section{Introduction}\label{sec:introduction}%
  \global\everypar=\everypar}}%
  \vspace{-1\baselineskip}\vspace{-\parskip}\par
\else
  \section{Introduction}\label{sec:introduction}\par
\fi

In recent years, data analysis methods are increasingly being used in order to automate the extraction of relevant information from data collections. However, most data analysis algorithms suffer from overfitting and numerical problems when patterns have more than a few dimensions. In order to appropriately apply these \cblue{data analysis methods}, a feature extraction pre-processing stage is required, allowing to remove collinearity among variables and to reduce data dimensionality. For this reason, feature extraction techniques, and in particular Multivariate Analysis (MVA) methods \cite{Mardia1980,Arenas13}, have been successfully applied in many different applications of machine learning, such as biomedical engineering \cite{Gerven12,Hansen07}, remote sensing \cite{Arenas08,Arenasbook}, or chemometrics \cite{Barker03}.

MVA aggregates a family of methods that build a new set of features by  extracting highly correlated projections of data representations in input and target spaces. Well-known representatives of these methods are Principal Component Analysis (PCA)  \cite{pearson1901pca}, Partial Least Squares (PLS) approaches \cite{wold1966nipals1}, or  Canonical Correlation Analysis (CCA) \cite{hotelling1936cca}. PCA creates a new data representation space by finding the directions of largest input variance, providing an optimal set of features in terms of mean-squared reconstruction error. Unlike other MVA methods, PCA works in an unsupervised manner, i.e., it only considers the input data and does not take into account any \cblue{knowledge of the target space}. The Partial Least Squares (PLS) approach refers to a family of techniques which, in its general form, project both input and target variables to a new space, generating a set of projected features known as latent variables. The criterion used to extract these latent variables varies within the approach, but the general objective is to maximize the covariance of the two projected expressions. In CCA the aim is to find linear projections of the input and target data to maximize correlation between the projected data.

In this paper, we focus on a fourth MVA method known as Orthonormalized PLS (OPLS) \cite{worsley1998mvlm}, also known as semipenalized CCA \cite{Barker03}, multilinear regression (MLR) \cite{Borga97}, or reduced-rank regression \cite{reinsel98}. OPLS is known to be optimal in the mean square error sense for performing multilinear regression \cite{Roweis99,Arenas07NIPS}; thus, it is very competitive as a pre-processing step in classification and regression problems  \cite{Arenas08,Arenas07NIPS,Dhanjal09}. Several works have also tried to establish the connections between OPLS and other MVA and discriminative methods (see, e.g., \cite{reinsel98,Sun09} which proved that OPLS and CCA are the same solution in a balanced classification task if the target matrix is binary coded, or \cite{Torre2012} that proposes a generalized framework for component analysis).

\begin{figure*}[t]
\centering
\includegraphics[scale=.55]{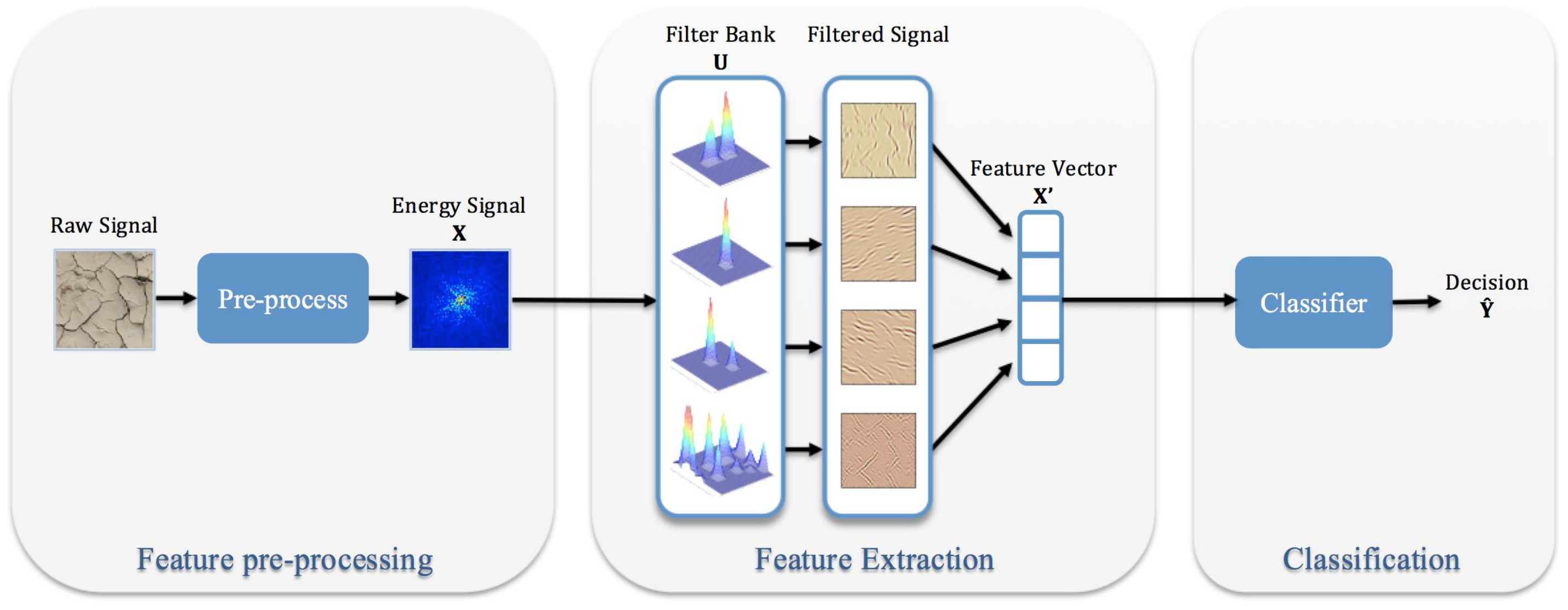}
\caption{Texture image recognition scheme from the raw image to the decision. The image is primarily processed to obtain a 2-D frequency representation, which is then passed through the filter bank so that each extracted feature summarizes the energy contained in a certain frequency range. Finally, classification is done based on the extracted features.}
\label{fig:scheme}
\end{figure*}

In this paper, we consider extensions of OPLS that favor the interpretability of the extracted features. In particular, when energy or spectral signals are involved, positive constraints should be imposed on the projection vectors, so that the extracted features can be interpreted as the energy contained in an equalized frequency band, and the projection vectors themselves can be seen as a sort of filter bank. This interpretation is useful, e.g., in audio or image data applications, where data processing is usually carried out in \cblue{the} frequency domain. Note that, in this paper, we pursue interpretable solutions, considering interpretability as the machine capability of \cblue{revealing to the expert the reasons why the solution has been reached.}

We can find in the recent machine learning literature other algorithms preserving the non-negativity of the solution. One of the most popular algorithms is Non-Negative Matrix Factorization (NMF) \cite{Lee1999} that has been applied for source separation \cite{kounades2016variational}, music transcription \cite{Smaragdis2003}, or spectral data analysis \cite{Pauca2006}, among others. 
Another perhaps less explored approach consists  of incorporating a non-negativity constraint in the solution of the MVA methods. For instance, non-negative PCA has been used for blind positive source separation \cite{Oja2003}; 
non-negative PLS (NPLS) for understanding Nuclear Magnetic Resonance (NMR) spectroscopy data \cite{Allen2013}; non-negative CCA (NCCA) for audiovisual source separation \cite{Sigg2007}; and the positive constrained OPLS algorithm for music instrument and genre classification \cite{ArenasPOPLS}.

In contrast to NMF approaches, an additional advantage of incorporating non-negativity constraints in MVA methods is the capability to obtain sparse solutions and, indirectly, automatic feature selection. This preference for sparsity has motivated that during the last years many methods incorporate $\ell_0$ and $\ell_1$ regularizations in their formulations, see among others sparse versions of PCA \cite{Zou06}, CCA \cite{hardoon2011sparse}, and OPLS \cite{Sergio13}. However, unlike the methods we consider in this paper, neither $\ell_0$ nor $\ell_1$ regularization alone enforce non-negative solutions.

As we have already mentioned, in this paper we will focus on the design of banks of filters that provide interpretable features for supervised problems. Figure \ref{fig:scheme} illustrates the whole process when dealing with images, and consists of three well-differentiated blocks: 1) a pre-processing step which converts raw data into a better-fitted data representation in the frequency domain (see Sections \ref{section:textures} and \ref{section:music_genre}); 2) a feature extraction step where the signal is passed through a bank of filters and, as a result, a feature vector $\xz'$ is obtained, with each of its components being the energy of the image in a certain frequency range; and 3) a classification stage where the feature vector $\xz'$ is used to discriminate the class associated to the image.

In most previous works that rely on systems similar to the one depicted in Figure \ref{fig:scheme}, it is the classifier block which is designed in a supervised manner, whereas the filter bank is typically built without any available label information. Instead, a sufficiently rich battery of general purpose filters is used (e.g. Gabor filters \cite{Turner1986,Fogel1989}) or expert knowledge is exploited. Unlike these schemes, in this paper we will present a non-negative constrained OPLS method that exploits the available labels to design a bank of filters particularly fitted to each supervised task. As we will see, this results in more discriminative filters that can achieve better recognition rates with a smaller number of features. In order to show the versatility of our methods, we review two completely different scenarios: texture classification as a visual application; and music genre classification as an audio application.

Among a large amount of visual tasks, the classification of images by their textures is an interesting application. Surprisingly, the non-negative methods referred above have not been applied here, perhaps due to the wide and successful use of \cblue{purposely desinged methods} feature extraction procedures. One of the most adopted techniques is Gabor Filtering \cblue{(GF)}, which was proposed for texture discrimination in \cite{Turner1986,Fogel1989} and is still used or even improved for efficient feature extraction \cite{Bianconi2007,Li2010}, and for rotation and scale-invariant texture classification 
\cite{han2007rotation}
. Local Binary Pattern (LBP) is also a successful technique used for texture classification \cite{Ojala2002} 
but it does not provide any sort of interpretation to the solution.

Regarding audio-based music classification applications \cite{Fu2011}, and in particular the music information retrieval (MIR) field, music genre classification has been an active research area in the last years. Despite the great variety of different approaches to solve this problem, feature extraction is a usual stage into these solutions \cite{Scaringella2006} and the use of sparse representations has been recently suggested as a way to boost performance \cite{Sturm2013cs}
. However, sparse features do not provide any sort of interpretability to the solution, which is a desirable property for understanding music structure. In order to provide this capability, non-negative features can be extracted, as it was proposed by \cite{ArenasPOPLS} and \cite{Mckinney2003}. 


\cblue{This paper proposes a general framework to obtain non-negative solutions for the desing of filter banks, providing solutions which not only perform well in classification tasks but also help to interpret the machine working}. For this purpose, starting from a non-negative OPLS formulation, we study four different approaches to solve the proposed \cblue{framework}. The performance of these approaches is compared with classical filter design methods for image and music classification (where their parameters need to be adjusted by an expert user) and, additionally, with some current state of art approaches based on deep learning.

The rest of the paper is organized as follows: In the next section, we propose different ways of designing filter banks in a supervised way as a feature extraction stage for audio or visual applications. In Sections \ref{section:textures} and \ref{section:music_genre}, we review the most frequently used methods in the visual and audio fields, focusing in particular on the texture and music genre classification problems. In order to prove the general applicability of our methods for applications when energy or spectral data are involved, we analyze the discriminative power of the extracted filter banks (i.e., non-negative features) on several texture and music genre databases in Section \ref{section:experiments}. Finally, Section \ref{section:conclusions} presents the main conclusions of our work.

\section{OPLS-based Methods for Supervised Design of Filter Banks}
\label{section:designFilterBanks}

In this section, we formulate several methods to design filter banks in a supervised learning scenario, where the goal is to learn relevant features from input data using a set of $N$ training data $\{\bar{\xz}_i,\bar{\yz}_i\}$, for $i = 1, \dots, N$, where $\bar{\xz}_i \in \Re^n$ and $\bar{\yz}_i \in \Re^m$ are considered as the input and target data vectors, respectively. Therefore, $n$ and $m$ denote the dimensions of the input and target spaces. In classification problems, $\bar{\yz}_i$ will be used to denote the class membership of the $i$th pattern, e.g., using 1-of-$m$ encoding \cite{Bishop95}. Furthermore, in this paper, we assume that all entries of $\bar{\xz}_i$ are non-negative, which is the case when dealing with applications where the input space consists of spectral features. For notational convenience, we define the input and target data matrices: $\XX = \left[\xz_1,\dots,\xz_N \right]$ and $\YY = \left[\yz_1,\dots,\yz_N \right]$, where $\xz_i$ and $\yz_i$ are centered versions of $\bar{\xz}_i$ and $\bar{\yz}_i$, i.e.,
\begin{equation*}
\label{eq:centeredData}
\xz_i=\bar{\xz}_i-\muz_x, \qquad \yz_i=\bar{\yz}_i-\muz_y,
\end{equation*}
where $\muz_x$ and $\muz_y$ are the means of the input and target data. Sample estimations of the input and target data covariance matrices, as well as of their cross-covariance matrix, can then be calculated as $\CC_{\XX\XX} = \XX\XX^\top$, $\CC_{\YY\YY} = \YY\YY^\top$ and $\CC_{\XX\YY} = \XX\YY^\top$, where we have neglected the scaling factor $\frac{1}{N}$, and superscript $^\top$ denotes vector or matrix transposition.

We denote the desired filter bank as $\UU = [\uz_1,\dots,\uz_{n_f}]$\cblue{$\in \Re^{n\times n_f}$}, where each of its columns can be seen as the response of one of the $n_f$ filters in the bank and $\xz_i'=\UU^\top\xz_i$ represent the extracted features corresponding to pattern $\xz_i$.

When the input data are (positive) spectral features, we can consider $\UU$ as a frequency filter bank whose coefficients are going to be forced to be non-negative, and $\xz_i'$ can be interpreted as the non-negative output of each of the filters in the bank. However, when matrix $\XX$ is centered, $\XX'=\UU^\top\XX$ can also be seen as projections of the centered input data, but its entries are no longer guaranteed to be non-negative. Nevertheless, data centering does not affect the design of the filter bank and is recommended for learning purposes if regression procedures are involved \cite{Shawe04}. In fact, it is quite straightforward to illustrate the irrelevance of the centering operation with respect to the extracted features since
\begin{equation*}
\label{eq:centeredInputs}
\UU^\top\bar{\xz}_i=\UU^\top\xz_i-\UU^\top\muz_x=\xz_i'-\muz_x',
\end{equation*}
where $\muz_x'$ is the mean of the filtered data. Therefore, the interpretation of the filter bank remains valid when working with centered data, and the optimization problems become numerically more stable \cblue{\cite{Shawe04}}.

We use OPLS as a starting point in order to design the filter bank. OPLS is optimal in the Mean Square Error (MSE) sense \cite{Roweis99}, i.e. to find the projection vectors so that the projected data best approximate the target data minimizing $||\YY-\WW\UU^\top\XX||^2_F$, where subscript $F$ refers to the Frobenius norm of a matrix and $\WW$\cblue{$\in \Re^{m\times n_f}$} is the regression matrix. For this reason, OPLS usually outperforms all other traditional Multivariate Analysis (MVA) algorithms in both regression and classification tasks \cite{Arenas13}. However, to enforce non-negative filter coefficients and allow more interpretable solutions, we add a non-negativity constraint to the OPLS objective function. Hence, the minimization problem that we propose for the design of the filter bank is:
\begin{equation}
\label{Loss_function}
\begin{array}{ll}
\displaystyle\min_{\UU,\WW} & \|\YY  - \WW \UU^\top \XX \|_F^2 \\ 
\text{s.t.:~} & \UU \geq \0z
\end{array}
\end{equation}
where $\UU\geq\0z$ indicates that all elements of matrix $\UU$ have to be non-negative.

In this paper we propose four different algorithms to solve \eqref{Loss_function}:
\begin{enumerate}
\item Non-negative OPLS (NOPLS): based on alternating two coupled convex problems (i.e., solving for $\UU$ and $\WW$ iteratively) that we proposed in \cite{Sergio15} and \cite{Sergio16}. Here, the method is adapted to include non-negative constraints.
\item NOPLS with Procrustes approach (P-NOPLS): similar to NOPLS, but solving the iterations for $\WW$ using the ``Orthogonal Procrustes problem'' \cite{Schonemann1966}. Here, we adapt this technique as another alternative to solve \eqref{Loss_function}.
\item Deflated NOPLS: the sequential implementation of NOPLS using deflation. 
\item NMF-like OPLS (NMF-OPLS): that can be considered as a supervised version of the NMF problem. 
\end{enumerate}

Note that although all algorithms try to solve the same optimization problem, they will in general converge to different solutions. In the rest of this section we will introduce these algorithms, and compare their results in the experimental section. For completeness we will also consider a fifth method known as positive constrained OPLS (POPLS). This method was proposed in \cite{ArenasPOPLS} and relies on Quadratic Programming (QP) for solving \eqref{Loss_function}.

\subsection{Non-negative OPLS (NOPLS)}
\label{subsection_NOPLS}
The algorithm proposed in this subsection is based on the sparse OPLS (SOPLS) method proposed in \cite{Sergio13}. In that paper, an efficient solution of OPLS, also known as reduced rank regression problem in the statistics literature \cite{reinsel98}, was used in order to propose a feature extraction framework, which allows to easily include any constraints to the projection\footnote{Note that $\UU$ is not a projection operator in a rigorous mathematical sense, since it maps data from $\Re^n$ to $\Re^{n_f}$, and therefore does not satisfy the idempotent property of projection operators. However, the columns of $\UU$ span the subspace of $\Re^n$ where the data are projected, and it is in this sense that we refer to $\UU$ and $\uz_i$ as projection matrix and vector respectively, and to $\XX'$ as projected data. This nomenclature has been widely used in the machine learning field, particularly in works dealing with feature extraction methods.} matrix $\UU$. The algorithm consists of the iterative application of two coupled steps: a constrained least squares regression problem, and a standard eigenvalue problem with the same complexity as the target dimensionality. Note that in audio or visual applications, the target dimensionality is usually much smaller than the input dimensionality (i.e., $m \ll n$) and, thus, this latter step is normally quite efficient.

In this subsection, we propose to replace the sparsity constraint that led to SOPLS by the desired non-negativity constraint. Therefore, according the same arguments of \cite{Sergio13}, we propose the following iterative procedure to solve the objective function \eqref{Loss_function}:

\begin{enumerate}
\item[1)] $\WW-$step: For fixed $\UU$, minimize \eqref{Loss_function} with respect to $\WW$, subject to $\WW^\top \WW = \II$.

The solution of this problem is given by the eigenvalue decomposition
\begin{equation}
\CC_{\XX'\YY}^\top \CC_{\XX'\YY} \WW = \WW \Lambdaz, \label{Wstep_NOPLS}
\end{equation}
where $\CC_{\XX'\YY} = \UU^\top\CC_{\XX\YY}$ \cblue{and $\Lambdaz$ is a diagonal matrix with all the eigenvalues sorted in descending order on its diagonal}. Note that the dimension of the matrix that needs to be analyzed is $m$, which makes up a \cblue{computationally} efficient step in the common case of $m<n$. 


\item[2)] $\UU-$step: For fixed $\WW$, minimize \eqref{Loss_function} with respect to $\UU$ only.

We refer the reader to \cite{van2004fast} and \cite{Kim2008} for good summaries on optimization methods that can be used to solve this non-negative least squares (NNLS) problem. In the experiments section, we will use the MATLAB implementation of block principal pivoting algorithm provided by \cite{Kim2008}\footnote{Code is available at \url{http://www.cc.gatech.edu/~hpark/software/nmf_bpas.zip}}. In case of adding also an $\ell_1$-regularization term to \eqref{Loss_function} (i.e., $\lambda ||\UU||_1$), the Monotone Incremental Forward Stagewise Regression (MIFSR) algorithm proposed in \cite{Hastie2007forward} with the modifications introduced in \cite{Sigg2007} can be used.
\end{enumerate}

The NOPLS method consists  of the iterative application of the $\WW$ and $\UU-$steps until some convergence criterion is reached. Preliminary experiments have showed that initialization of the algorithm is not critical, and we simply initialize $\UU$ for the first iteration as \cblue{a matrix whose elements $u_{ij}$ are defined by the Kronecker delta $\delta_{ij}$}. As a stopping mechanism, we use $\Trc\{\Lambdaz^{(k)} - \Lambdaz^{(k-1)}\} \leq \delta$, where the superscripts denote the iteration index and $\delta$ is a small constant. In plain words, the algorithm stops when the difference between the eigenvalues of the $\WW-$step of two consecutive iterations is smaller than a prefixed constant $\delta$.

\subsection{NOPLS with Procrustes approach (P-NOPLS)}
\label{subsection:P-NOPLS}
Our second proposed method consists  of the modification of the $\WW-$step of NOPLS applying the solution to the orthogonal Procrustes problem. This approach was used by \cite{Zou06} and \cite{Gerven12} to derive sparse solutions of PCA and OPLS respectively. As we explained above, when fixing the projection matrix $\UU$ the $\WW-$step of the algorithm becomes:
\begin{equation}
\label{Procrustes_problem}
\begin{array}{ll}
\displaystyle\min_{\WW} & \|\YY  - \WW \XX' \|_F^2 \\ 
\text{s.t.:~} & \WW^\top\WW=\II.
\end{array}
\end{equation}
In \cite{Schonemann1966}, this problem is called an ``Orthogonal Procrustes problem'' and its solution is defined as
\begin{equation}
\label{Procrustes_solution}
\WW_{\text{PROCRUSTES}} = \mathbf{Q}\mathbf{P}^\top,
\end{equation}
given the singular value decomposition $\CC_{\XX'\YY} = \mathbf{P}\mathbf{D}\mathbf{Q}^\top$, where $\mathbf{D}$ is a diagonal matrix containing the singular values, and $\mathbf{P}$ and $\mathbf{Q}$ contain the left and right singular vectors, respectively.

Since the solution of \eqref{Wstep_NOPLS} gives $\WW_{\text{NOPLS}}=\mathbf{Q}$, we can see that P-NOPLS results just in a rotated version of this matrix during the $\WW-$step. However, we note that:
\begin{itemize}
\item The rotation process affects the relevance and ordering of the extracted features. For NOPLS we can affirm that the features/filter banks are sorted according to their relevance, i.e., \cblue{the} first filter captures the maximum possible information with just one filter with respect to criterion \eqref{Loss_function}, and so on. The rotation process prevents us from stating this same claim for P-NOPLS.
\item In \cite{Sergio13}, we showed that the Procrustes solution is very sensitive to initialization and that for some initializations the algorithm may fail to converge.
\end{itemize}

The two arguments above justify our preference for NOPLS over the P-NOPLS solution. Nevertheless, P-NOPLS has also been included in the experiments for the sake of comparison and completeness.



\subsection{Sequential implementation of NOPLS using deflation (defNOPLS)}
\label{subsection:deflatedNOPLS}
In this section, we describe a sequential algorithm that implements the non-negative OPLS scheme introduced in Subsection \ref{subsection_NOPLS}. This sequential algorithm consists of the following two steps: 1) Extraction of the projection vector $\uz_j$ which represents the frequency response of the next filter to be included in the bank, and 2) application of a deflation procedure to eliminate the influence of the $j^{th}$ eigenvector by annihilating the associated eigenvalue. \cblue{Note that the deflation process allows to obtain a new data space where the resulting matrix does not depend on the space defined by the $j^{th}$ eigenvector, so that we can again apply the first step and calculate the next most important eigenvector.} 
 These steps are repeated for $j=1,\dots,n_f$ until the desired number of filters or features is reached.

The design of the next filter consists  of the extraction of a pair of vectors $\{\uz_j,\wz_j\}$ which are optimal with respect to \eqref{Loss_function}. This can be done by iterating the $\WW$ and $\UU-$steps we have described for the NOPLS algorithm. Since in this case we are solving a unidimensional problem at each step, the solution to the $\WW-$step simplifies to
\begin{equation}
\label{eq_wzi}
\wz_j = \frac{\CC_{\xx'\YY}^\top}{\|\CC_{\xx'\YY}\|_{2}},
\end{equation}
where $\CC_{\xx'\YY} = \uz_j^\top \CC_{\XX\YY}$.


Throughout this paper, we will deflate the cross-covariance matrix by means of the Schur Complement deflation, which is one of the deflation methods proposed by \cite{Mackey2008}, and can be applied to any of a number of constrained MVA methods or eigendecomposition-based problems, including this non-negative constrained OPLS solution. This deflation procedure avoids the reappearance of $\uz_j$ as a component of future ``pseudo-eigenvectors''\footnote{In \cite{Mackey2008}, the ``pseudo-eigenvector'' term is used to differentiate the true eigenvector obtained from the original objective function (i.e. without any constraint on the eigenvector) and the solution obtained when constraints are applied.}. 
Since, in our case, we have to deal with a supervised problem, the Schur Complement deflation can be written as
\begin{equation}
\label{deflation}
\CC_{\XX\YY} \leftarrow \CC_{\XX\YY} \left[ \II - \frac{\CC_{\XX\YY}^\top \uz_j \uz_j^\top \CC_{\XX\YY}}{\uz_j^\top \CC_{\XX\YY} \CC_{\XX\YY}^\top \uz_j} \right],
\end{equation}
where $\leftarrow$ represents an update of the left matrix, and it renders $u_j$ left orthogonal to $\CC_{\XX\YY}$ only, since after deflation it holds that
\begin{multline*}
\uz_j^\top\CC_{\XX\YY} \left[ \II - \frac{\CC_{\XX\YY}^\top \uz_j \uz_j^\top \CC_{\XX\YY}}{\uz_j^\top \CC_{\XX\YY} \CC_{\XX\YY}^\top \uz_j} \right] \\= \uz_j^\top\CC_{\XX\YY} - \frac{ \uz_j^\top\CC_{\XX\YY}\CC_{\XX\YY}^\top \uz_j \uz_j^\top \CC_{\XX\YY}}{\uz_j^\top \CC_{\XX\YY} \CC_{\XX\YY}^\top \uz_j} = \0z
\end{multline*}

Table \ref{SOPLS_pseudocode} provides the pseudocode for the sequential algorithm that we have just described. Note that, in the table, subscript $j$ is used to index the projection vectors (i.e., $j=1,\dots,n_f$), whereas superscript $k$ indexes the iterative application of $\WW-$ and $\UU-$steps that are needed to converge to each projection vector. Different convergence criteria can be used at Step 2.2.3 of the algorithm. In the experimental section we will monitor the cosine distance
\begin{equation}
\label{eq:dirCosine_linear}
d_{\cos}\left(\uz_j^{(k)},\uz_j^{(k-1)}\right) = \frac{\uz_j^{(k)\top}\uz_j^{(k-1)}}{\|\uz_j^{(k)}\|_{2} \|\uz_j^{(k-1)}\|_{2}},
\end{equation}
and use as a stopping criterion $d_{\cos}\left(\uz_j^{(k)},\uz_j^{(k-1)}\right) > 1-\delta$, where $\delta$ is a tolerance parameter. Other possibilities would consist in monitoring the cosine distance between the regression coefficient vectors, or the eigenvalue of the $\WW-$step.

\begin{table}[ht]
\renewcommand{\arraystretch}{1.3}
\caption{Pseudocode for the sequential NOPLS using deflation.}
\centering
\label{SOPLS_pseudocode}
\begin{tabular}{l}
\toprule
1.-  Inputs: centered matrices $\mathbf{X}$ and $\mathbf{Y}$, $n_f$.\\
2.-  For $j=1,\dots,n_f$ \\
$\quad$ 2.1.- Initialize $\uz_j^{(1)} = \delta_j$ $~^\ddag$. \\
$\quad$ 2.2.- For $k = 1, 2, \dots$ \\
$\quad$ $\quad$ 2.2.1.- Update $\wz_j^{(k)}$ using \eqref{eq_wzi}.\\
$\quad$ $\quad$ 2.2.2.- Update $\uz_j^{(k)}$ by solving the unidimensional version\\$\quad$ $\quad$ ~~~~~~~ of the NNLS problem \eqref{Loss_function} subject to $\uz_j^{(k)} \geq 0$.\\
$\quad$ $\quad$ 2.2.3.- If convergence criterion is reached, provide output\\$\quad$ $\quad$ ~~~~~~~ values with $\{\uz_j,\wz_j\}$, otherwise back to 2.2. \\
$\quad$ 2.3.- Deflate cross-covariance matrix using \eqref{deflation}. \\
3.- Output: $\UU=[\uz_1,\dots,\uz_{n_f}]$.\\
\bottomrule
$~^\ddag$ $\delta_j$ is defined as a vector with its $j^{th}$ component equal to 1, and all\\ other components equal to 0.
\end{tabular} 
\end{table}

\subsection{NMF-like OPLS (NMF-OPLS)}

In this subsection, we solve problem \eqref{Loss_function} using an NMF approach, particularly we recur to the Multiplicative Updating (MU) rule proposed by \cite{Seung2001}, which is maybe the most popular NMF algorithm. 
Moreover, the loss function of the Projected-NMF algorithm proposed in \cite{Yuan2005} and some relationships among several versions \cite{Choi2008} can be useful to see the similarities between NMF and the supervised version we propose here.

Unlike previous methods, the NMF methods require non-negative values both for $\XX$ and $\YY$ (i.e., $\XX\geq \0z$ and $\YY\geq \0z$) and, thus, the additional constraint $\WW\geq \0z$ needs to be considered. Since certain data will take negative values due to the centering operation, we will use the original data set $\bar{\XX}$ and $\bar{\YY}$, which is non-negative.

The loss function to be minimized is also given by \eqref{Loss_function}, although in this case the constraint $\WW \geq \0z$ will be added:
\begin{equation}
\label{eq:NMF_lossFunction}
\begin{array}{ll}
\displaystyle\min_{\UU,\WW} & \|\bar{\YY}  - \WW \UU^\top \bar{\XX} \|_F^2 \\ 
\text{s.t.:~} & \UU \geq \0z, ~\WW \geq \0z.
\end{array}
\end{equation}
In order to ease the derivation of the current proposal, we rewrite the objective function of \eqref{Loss_function} in terms of the trace operator ($||\AAA||^2_F=\Trc\{\AAA\AAA^\top\}$):
\begin{multline}
\label{eq:NMF-OPLS_cost_traces}
{\cal L}(\WW,\UU) = \Trc\{\CC_{\bar{\YY}\bar{\YY}}\} - 2 \Trc\{\WW^\top \CC_{\bar{\XX}\bar{\YY}}^\top\UU\} \\+ \Trc\{\UU^\top \CC_{\bar{\XX}\bar{\XX}} \UU \WW^\top \WW\}.
\end{multline}

As a summary of the MU rule, let us suppose that the gradient of \eqref{eq:NMF-OPLS_cost_traces} with respect to $\UU$ or $\WW$ can be decomposed as
$$\partial{\cal{L}} = \partial{\cal{L}}^+ - \partial{\cal{L}}^-,$$
where $\partial{\cal{L}}^+ \geq 0$ and $\partial{\cal{L}}^- \geq 0$.
Then, the element-wise updating rule follows as \cite{Choi2008}:
\begin{equation}
\label{eq:MU_rule}
\Psi \leftarrow \Psi\circ\frac{\partial{\cal{L}}^-}{\partial{\cal{L}}^+},
\end{equation}
where $\circ$ denotes the Hadamard (element-wise) product, $\frac{\AAA}{\BBB}$ represents the element-wise division, i.e., $\left[\frac{\AAA}{\BBB}\right]_{ij} = \frac{A_{ij}}{B_{ij}}$ (for the $i^{th}$ row and the $j^{th}$ column), and $\Psi$ is the matrix that needs to be updated. Note that this update keeps the non-negativity of the solution $\Psi$ at every step.

To apply the MU rule in our case, we obtain first derivatives of \eqref{eq:NMF-OPLS_cost_traces} with respect to $\UU$
$$\frac{\partial {\cal{L}}(\UU,\WW)}{\partial \UU} = - 2 \CC_{\bar{\XX}\bar{\YY}} \WW + 2 \CC_{\bar{\XX}\bar{\XX}} \UU \WW^\top \WW,$$
that, considering that all involved matrices are non-negative, allows us to identify
$$\partial{\cal{L}}^+_\UU = 2\CC_{\bar{\XX}\bar{\XX}} \UU \WW^\top \WW,\qquad \partial{\cal{L}}^-_\UU = 2\CC_{\bar{\XX}\bar{\YY}} \WW.$$
Similarly, first derivatives of \eqref{eq:NMF-OPLS_cost_traces} with respect to $\WW$ are
$$\frac{\partial {\cal{L}}(\UU,\WW)}{\partial \WW} = - 2 \CC_{\bar{\XX}\bar{\YY}}^\top \UU + 2 \WW \UU^\top\CC_{\bar{\XX}\bar{\XX}} \UU,$$
so that we can identify
$$\partial{\cal{L}}^+_\WW = 2\WW \UU^\top\CC_{\bar{\XX}\bar{\XX}} \UU,\qquad \partial{\cal{L}}^-_\WW = 2\CC_{\bar{\XX}\bar{\YY}}^\top \UU.$$

Therefore, following equation \eqref{eq:MU_rule}, the MU rules for updating $\UU$ and $\WW$, which constitute the core of the NMF-OPLS method, are given by
$$\WW \leftarrow \WW\circ\frac{\CC_{\bar{\XX}\bar{\YY}}^\top \UU}{\WW \UU^\top\CC_{\bar{\XX}\bar{\XX}} \UU},
\qquad \UU \leftarrow \UU\circ\frac{\CC_{\bar{\XX}\bar{\YY}}\WW}{\CC_{\bar{\XX}\bar{\XX}}\UU\WW^\top\WW}.$$

As it is expected in NMF algorithms, preliminary experiments showed us that the algorithm initialization is critical. The nonnegative decomposition approximation method\footnote{We have used the NNDSVDa version of the algorithm in \cite{Boutsidis2008}, as well as the implementation provided by its authors.} (NNDSVD) proposed by \cite{Boutsidis2008} provides a good starting point for NMF algorithms, so we use it over the $\CC_{\bar{\XX}\bar{\YY}}$ matrix, since we deal with a supervised scheme. Therefore, we initialize the $\UU$ and $\WW$ matrices as the left and right-decomposition matrices respectively by the NNDSVD approach: $\CC_{\bar{\XX}\bar{\YY}} \sim \UU\WW^\top$.

Non-negative constraints usually force a large number of zeros in the solution matrix, often causing numerical problems that make the MU update get stuck earlier than desired. In \cite{Gillis2012}, it was proved that a slight improvement is reached substituting zeros by a small constant (e.g., $\epsilon=10^{-16}$). Thus, the improved MU rules are given by
\begin{eqnarray}
\WW & \leftarrow & \max\left(\epsilon,\WW\circ\frac{\CC_{\bar{\XX}\bar{\YY}}^\top \UU}{\WW \UU^\top\CC_{\bar{\XX}\bar{\XX}} \UU}\right),\label{W_update}\\
\UU & \leftarrow & \max\left(\epsilon,\UU\circ\frac{\CC_{\bar{\XX}\bar{\YY}}\WW}{\CC_{\bar{\XX}\bar{\XX}}\UU\WW^\top\WW}\right).\label{U_update}
\end{eqnarray}

Furthermore, 
\cblue{a normalization step in each MU update iteration in order to \cblue{facilitate} the convergence can be included in NMF algorithms \cite{Lin07}}. In our case, this step is also applied and the $\UU$ and $\WW$ matrices are normalized with \cblue{their} Frobenius \cblue{norms} respectively.

Table \ref{NMF-OPLS_pseudocode} provides the pseudocode for the NMF-OPLS algorithm that we have just described. Different convergence criteria can be used at Step 2.2.4 of the algorithm. In the experimental section we use $||\UU^{(k)} - \UU^{(k-1)}||_F \leq \delta$ as a stopping mechanism, where the superscripts denote the iteration index and $\delta$ is a small constant. Then, the algorithm is stopped when the solutions achieved in two consecutive iterations differ less than a small threshold.

\begin{table}[ht]
\renewcommand{\arraystretch}{1.3}
\caption{NMF-OPLS pseudocode.}
\centering
\label{NMF-OPLS_pseudocode}
\begin{tabular}{ll}
\toprule
1.-  Inputs: positive matrices $\bar{\XX}$ and $\bar{\YY}$.\\
$\quad$ 2.1.- Initialize $\WW^{(1)}$ and $\UU^{(1)}$ with NNDSVD algorithm.\\
$\quad$ 2.2.- For $k = 1, 2, \dots$ \\
$\quad$ $\quad$ 2.2.1.- Update $\WW^{(k)}$ using \eqref{W_update}.\\
$\quad$ $\quad$ 2.2.2.- Update $\UU^{(k)}$ using \eqref{U_update}.\\
$\quad$ $\quad$ 2.2.3.- Normalize $\WW^{(k)}$ and $\UU^{(k)}$.\\
$\quad$ $\quad$ 2.2.4.- If convergence criterion is reached, go to 3. \\
3.- Outputs: $\UU$, $\WW$.\\
\bottomrule
\end{tabular} 
\end{table}

A few considerations are in order with respect to the algorithm we have just described. First, the main advantage of the MU rule is its simplicity and ease of implementation; however, slow convergence is frequently observed \cite{Kim2008}. Second, the application of NMF-OPLS requires that the number of filters in the bank ($n_f$) is selected a priori, and 
sequential implementations are suboptimal since the positive constraint prevents the deflation procedure from completely removing the contribution of the previous projections.
Finally, unlike NOPLS, the NMF-based implementation does not guarantee the filters of the bank (i.e., the columns of $\UU$) to be sorted according to their relevance.

\subsection{Positive Constrained OPLS (POPLS)}
For completeness, this subsection describes the algorithm proposed in \cite{ArenasPOPLS} to solve \eqref{Loss_function}. In this case, matrix $\WW$ is not explicitly calculated since the trick here is to express $\WW$ in terms of $\UU$ and to introduce it into the functional in \eqref{Loss_function} to obtain an optimization problem involving $\UU$ only.

The optimal regression matrix can be calculated minimizing \eqref{Loss_function} with respect to $\WW$ only, being the solution 
$\WW=\CC_{\XX\YY}^\top \UU\left(\UU^\top\CC_{\XX\XX} \UU\right)^{-1}.$
Introducing this result into \eqref{Loss_function}, we can rewrite the objective as a function of $\UU$ only
\begin{align*}
{\cal L}(\UU) &= \| \YY - \CC_{\XX\YY}^\top \UU\left(\UU^\top\CC_{\XX\XX} \UU\right)^{-1} \UU^\top \XX \|_F^2\\
&= \Trc\{\CC_{\YY\YY}\} - \Trc\{\left(\UU^\top\CC_{\XX\XX} \UU\right)^{-1} \UU^\top \CC_{\XX\YY}\CC_{\XX\YY}^\top\UU\}.
\end{align*}
Thus, we arrive to the following optimization problem
\begin{equation}
\label{POPLS_problem}
\begin{array}{ll}
\displaystyle\max_{\UU} & \Trc\{\left(\UU^\top\CC_{\XX\XX} \UU\right)^{-1} \UU^\top \CC_{\XX\YY}\CC_{\XX\YY}^\top\UU\} \\ 
\text{s.t.:~} & \UU\geq0, ~\UU^\top\UU=\II,
\end{array}
\end{equation}
where the last constraint is added in order to obtain one of the infinite solutions of objective function in \eqref{POPLS_problem}. Note that this constraint is different from that of the OPLS problem. However, this constraint was preferred in \cite{ArenasPOPLS} since it can be directly incorporated into the hyperspherical representation of the projection vectors, where each $\uz_j$ is represented by a radius $r_j$ and $n-1$ angles $\theta_j^{(s)}$, $s=1,\dots,n-1$. This way, the optimization can be solved with respect to $\theta_j^{(s)}$ for $r_j=1$, and constraints $0\leq\theta_j^{(s)}\leq \frac{\pi}{2}$ guarantee the non-negativity of the solution. This approach was followed in \cite{ArenasPOPLS} to solve the convergence problems of the \textit{fmincon} matlab function with the original representation used in \eqref{POPLS_problem}.

An inconvenience of this method is that the desired OPLS property $\UU^\top\CC_{\XX\XX}\UU=\II$ is not satisfied, implying that filters are not arranged according to their discriminative power. To correct this, in this paper we apply a sequential implementation using Schur Complement deflation. The resulting sequential POPLS algorithm is summarized in Table \ref{POPLS_pseudocode}.

\begin{table}[ht]
\renewcommand{\arraystretch}{1.3}
\caption{Pseudocode for POPLS with deflation procedure.}
\centering
\label{POPLS_pseudocode}
\begin{tabular}{l}
\toprule
1.-  Inputs: centered matrices $\mathbf{X}$ and $\mathbf{Y}$.\\
2.-  For $j=1,\dots,n_f$ \\
$\quad$ 2.1.- Update $\uz_j$ by solving the unidimensional version of \eqref{POPLS_problem}, i.e.,\\
$\quad$ $\quad$ ~~ $\displaystyle\max_{\uz_j} ~ \frac{\uz_j^\top \CC_{\XX\YY}\CC_{\XX\YY}^\top\uz_j}{\uz_j^\top\CC_{\XX\XX} \uz_j},$\\
$\quad$ $\quad$ ~~ subject to $\uz_j\geq 0$ and $\|\uz_j\|_{2}=1$.\\
$\quad$ 2.2.- Deflate cross-covariance matrix using \eqref{deflation}. \\
3.- Outputs: $\UU=[\uz_1,\dots,\uz_{n_f}]$.\\
\bottomrule
\end{tabular} 
\end{table}

\section{Texture Classification Application}
\label{section:textures}
To illustrate the advantages of OPLS methods for supervised filter design, we will consider texture recognition and music genre classification applications. For this reason, this section provides a brief summary of the necessary background on image processing required to understand the application of OPLS methods for texture recognition and the alternatives that are currently available in the literature. Similarly, Section \ref{section:music_genre} will deal with the music genre recognition problem.

Figure \ref{fig:scheme} describes all the stages that are typically encountered in texture recognition. Following these stages, the image is firstly preprocessed and, then, it is transformed to \cblue{the} frequency domain to facilitate the extraction of relevant features through filtering. Finally, a classifier is used to discriminate among the different textures.

A typical simple \cblue{pre-processing} step is to apply a two-dimensional Fast Fourier Transform to each image ($\xz$, if it is vectorized) which is usually transformed into gray scale if the raw image is in color. This allows the next stage to extract features ($\xz'$) in a frequency domain by means of a filter bank ($\UU$) as $$\xz'=\UU\xz.$$

One of the most popular feature extraction techniques for texture classification is \cblue{GF}. However, \cblue{GF} show a strong dependence on several parameters whose values may significantly affect the discriminative performances of the subsequent classifier. 
The effects of \cblue{GF} parameters on texture classification was comprehensively evaluated by \cite{Bianconi2007}. 

\begin{figure*}[!t]
\centering
\includegraphics[width=6in]{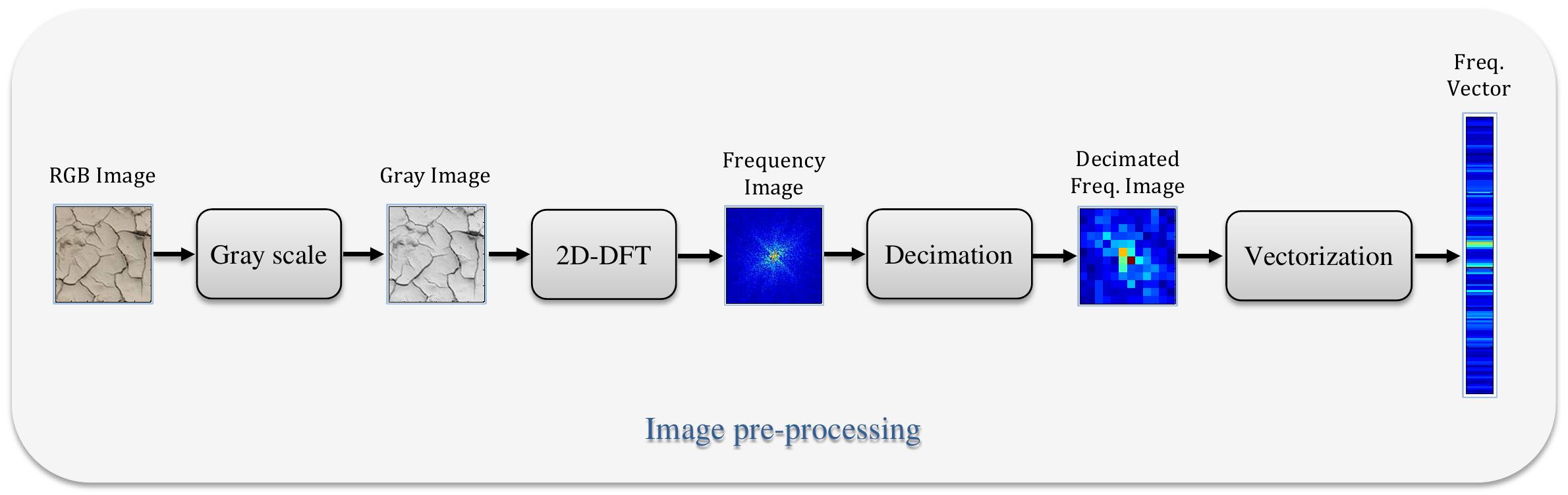}
\caption{Example of the pre-processing scheme applied to an image belonging to the earth class in the CGTextures dataset. The last two blocks are only included for our methods only.}
\label{Fig:imagePreprocessingStep}
\end{figure*}


One remarkable conclusion of \cite{Bianconi2007} is that smoothing parameters $\gamma$ and $\eta$ are important, whereas the number of frequencies and orientations has, in general, little effect on texture classification. This result contradicts the widely accepted belief that the parameters presenting a highest influence over the texture classification performance are related to the number of orientations ($n_O$), the number of frequencies ($n_F$), and the largest frequency of all filters.

As illustrated by the study in \cite{Bianconi2007}, the design of a GF bank can be very costly as a result of the validation process that is required to adjust the free parameters. Furthermore, the general shape of GFs is predefined a priori, and apart from their widespread use in texture classification, there are no guarantees that GFs are the most adequate selection for a particular task. In contrast to this, our proposed methods use available labels to build the bank of filters and do not assume any predefined shape for the frequency response of the filters. For this reason, we expect that they are able to extract more discriminative features for each particular supervised task.

To conclude the subsection, there are some practical considerations that need to be taken into account and imply differences between our schemes and the direct application of GF:
\begin{itemize}
\item  \cblue{GF} provides two features per filtered image: the mean ($\mu$) and the standard deviation ($\sigma$) of the filtered image. In contrast, our methods produce only one feature per filtered image.
\item In order to ease the operation of our algorithms, we decimate each frequency image by using the average energy over neighboring pixels of the image. This results in lower resolution of $\rho\times\rho$ and thus in a dimensionality reduction (of the vectorized frequency vector) to $n$ variables (being $n=\rho^2$). This pre-processing step is illustrated in Figure \ref{Fig:imagePreprocessingStep}. 
\end{itemize}

%

\section{Music Genre Classification}
\label{section:music_genre}

In this section we review the applicability of the OPLS-based schemes presented in this paper for music recognition applications. Although we consider here the particular case of music genre classification, the approach could be straightforwardly extended to other music information retrieval tasks. As before, the goal of the automatic design of the filter bank is to obtain good recognition rates, while at the same time extracting interpretable features.



\begin{figure}[!t]
\centering
\includegraphics[width=2.6in]{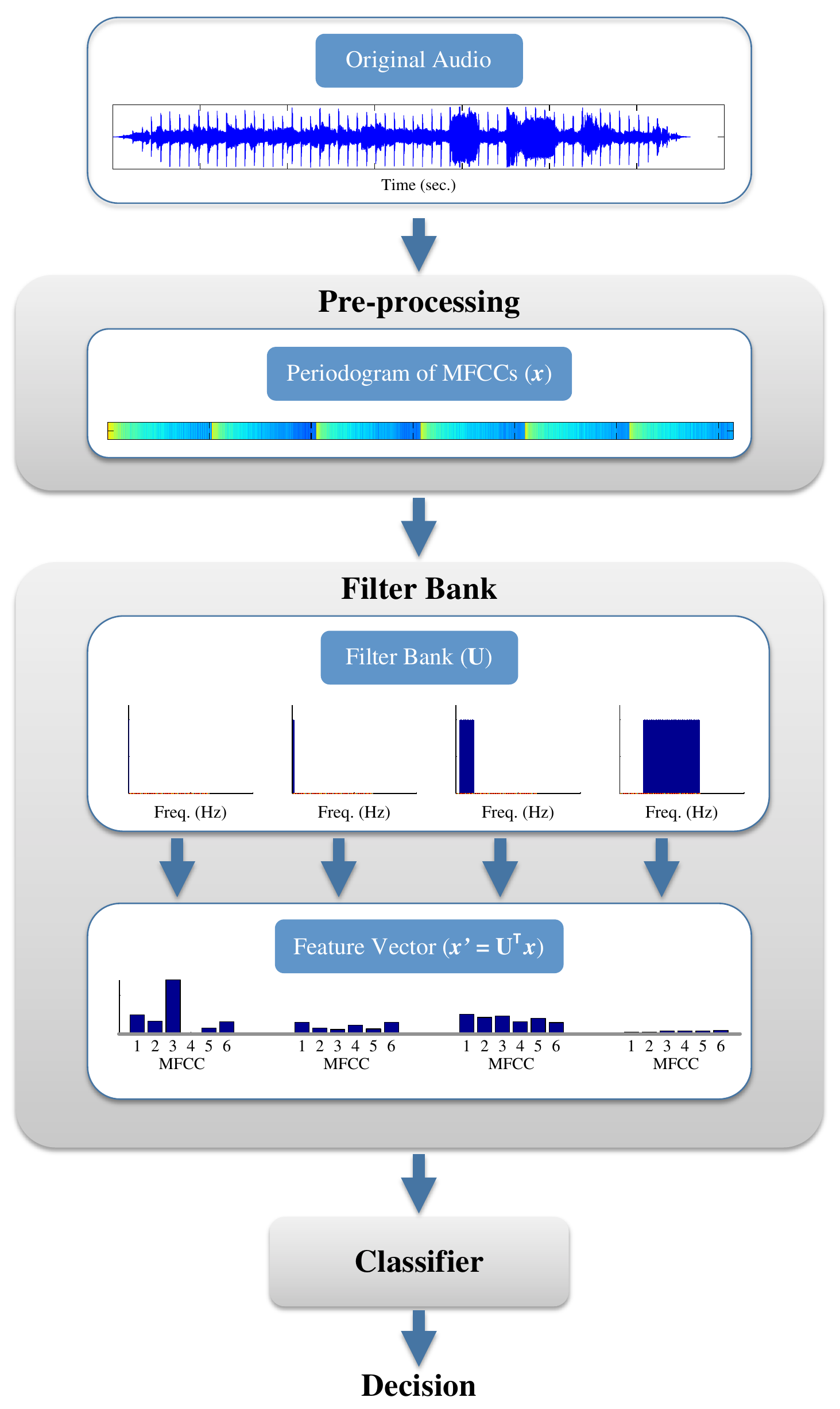}
\caption{Music genre classification scheme from the raw audio song to the decision. The audio clip is primarily processed to obtain a frequency representation, in this case a periodogram of the first 6 MFCCs. The periodograms are then passed through the filter bank, so that each extracted feature summarizes the energy contained in a certain frequency range. Finally, classification is carried out based on the extracted features.}
\label{Fig:GenreSchemeTotal}
\end{figure}

The whole music recognition application can be summarized into three well-differentiated blocks (see Figure \ref{Fig:GenreSchemeTotal})\cblue{, namely, the audio pre-processing step, a filter bank, and the classifier.}

\begin{figure}[!t]
\centering
\includegraphics[width=2.2in]{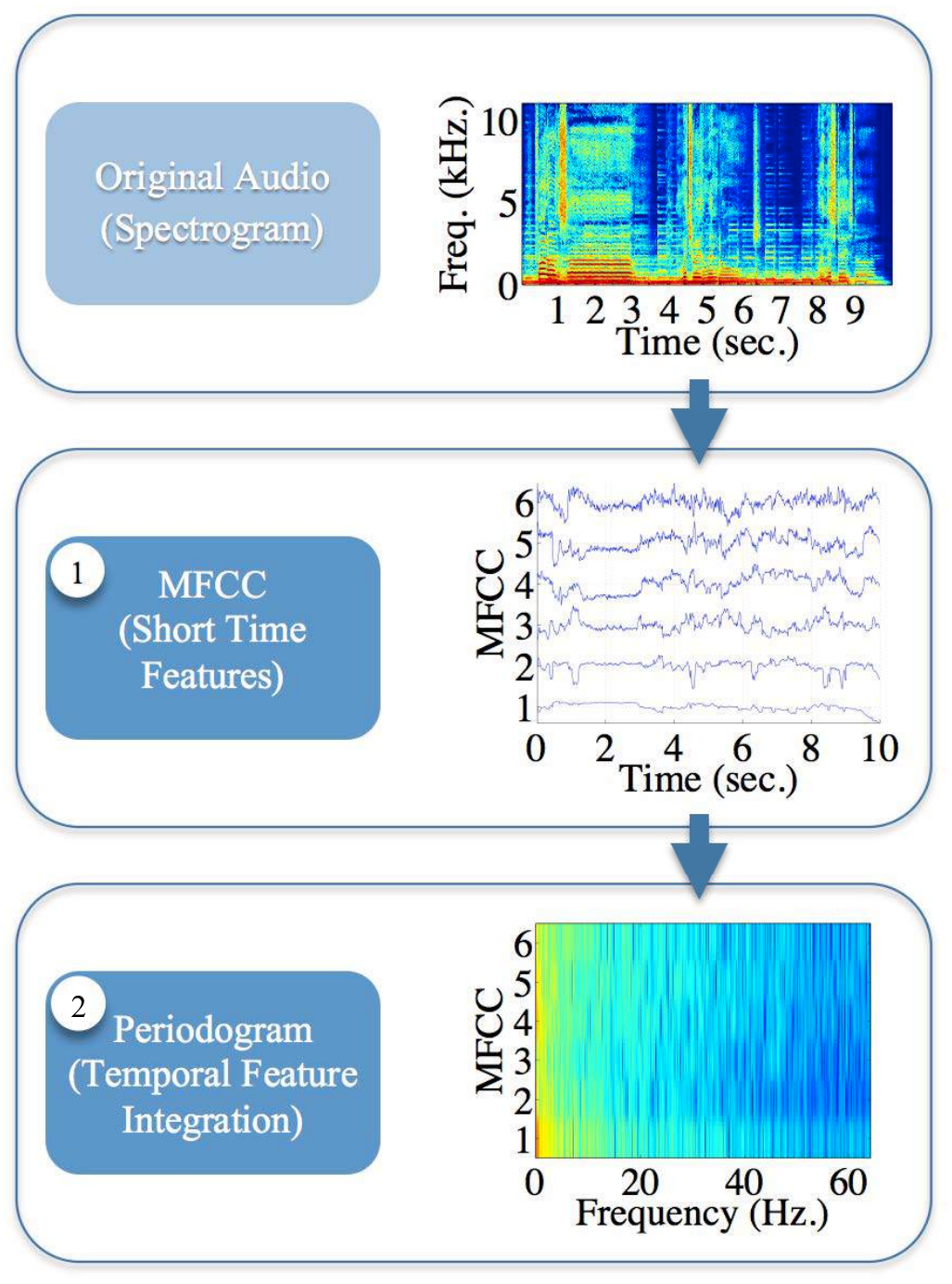}
\caption{Pre-processing step scheme of a ten second excerpt of the song ``Follow The Sun'' by ``Xavier Rudd''.}
\label{Fig:GenreScheme}
\end{figure}

The audio pre-processing step, which transforms raw audio signals into relevant information for the subsequent step, can be further subdivided into two stages (see Figure \ref{Fig:GenreScheme}): short-time feature extraction, which consists of features extracted in periods ranging from 5-100 ms where music signals are typically stationary, see e.g. \cite{Aucouturier2005}; and temporal feature integration, which is the process of combining all the feature vectors of a time frame into a single feature vector in order to capture the relevant temporal information of the frame.

In the following, we detail these two stages:

\begin{enumerate}
\item \textbf{Short-time features}: Mel Frequency Cepstral Coefficients (MFCC) have been selected as the short-time feature representation 
because of its widespread use and great success in several fields of MIR (see, e.g., \cite{Mckinney2003,Mandel2006}). The MFCCs are ranked in decreasing order of the richness of representation of the spectral envelope. Thus, the lower MFCCs contain information about the slow variations in the spectral envelope. 
In experiments, we use only the 6 first MFCCs (as proposed in \cite{ArenasPOPLS}) and, in order to minimize aliasing in the MFCCs, we apply a frame-size of 30 ms and a hop-size of 7.5 ms. 

\item \textbf{Temporal feature integration}: In order to capture the relevant temporal information in the frame, we first estimate the power spectrum of each MFCC using a periodogram as suggested in \cite{Mckinney2003}. Then, we concatenate these six energy features into a new single feature vector. There exist many other temporal feature integration methods, see \cite{Meng2007} for a good review on these techniques.
\end{enumerate}

Once the raw data are converted into a non-negative representation (i.e. the periodograms of the MFCCs, $\XX$), the following step relies on applying a filter bank, $\UU$, in order to extract the desired non-negative features, $\XX'=\UU^\top\XX$, which can be seen as the energy contained in certain frequency bands of each MFCC periodogram. Note that this filter bank $\UU$ is inherently non-negative as well, since it is applied directly on the estimated power spectrum (periodogram), $\xz_i'=\UU^\top\xz_i$, where $\xz_i$ is the periodogram of the $i^{th}$ MFCC coefficient, and $\xz_i'$ is the corresponding $i^{th}$ feature vector, which has as many components as the number of filters in the bank. These feature vectors will be introduced into the subsequent classifier.

In order to design the filter bank ($\UU$), there are two different alternatives: using expert knowledge, which is the most typically used approach; and the supervised approaches that we propose in the paper, which make use of the label information allowing the \cblue{learning of} filter banks for each recognition task. An example of the first alternative is the predefined Philips filter bank used in \cite{Mckinney2003}, where the authors suggest summarizing the power components in four frequency bands: 1) 0 Hz (DC component); 2) 0--2 Hz (beat rates); 3) 3--15 Hz (modulation energy, e.g., vibrato); 4) $>20$ Hz (associated to the perceptual roughness).  Therefore, for this particular filter bank, U is a matrix of size $D\times 4$, where $D = \frac{f_s}{2} + 1$ is the number of points of the periodogram and $f_s$ is the length of the MFCC series used to calculate the periodogram (measured in number of samples). In this paper, we will use $f_s = 256$. In Subsection \ref{subsection:genreClassification}, we compare our solutions with this fixed filter bank.



\section{Experiments}
\label{section:experiments}
%
In this section, we analyze the performance of the proposed set of supervised filter banks in two classification tasks: image texture recognition and  music genre classification. In order to evaluate the advantage of our proposals, we analyze their discriminative power and their interpretability capability in comparison to the well studied Gabor and Philips \cblue{Filter Banks}, which are \cblue{specifically} designed for the considered applications. Moreover, we compare the proposed methods against deep learning approaches, which are the \cblue{state-of-the-art} in \cblue{these} problems, specially in visual applications as texture classification. Therefore, we can evaluate if the interpretable solutions obtained with our methods can be competitive in terms of performance.

In order to evaluate the interpretability of the solutions, we need to define some objective criterion. Focusing in these multimedia applications, we consider that the interpretability of the solutions is higher when 1) we use a small number of filters ($n_f$), and 2) the number of non-zero energy bands in each feature is small. We denote as $\text{NZ}$ the rate of non-zero coefficients of the filters\cblue{, $\text{NZ}=\frac{k}{n_f n}$, where $k$ is the number of non-zero coefficients of the filters}. However, \cblue{given} that these two interpretability criteria are of different nature and are subject to a tradeoff (a similar performance can be obtained with fewer but less sparse features), it is difficult to combine them into a single measure of interpretability. Moreover, the preference of either criterion will depend on the user or the application. Due to these reasons, in this paper we will \cblue{report} these two terms separately, but we also propose and use a combined interpretability measure:
\begin{equation}
\label{eq:IM}
\text{IM} = -\log(\text{NZ})-\log(n_f/n_{ref}),
\end{equation}
where $n_{ref}$ is a reference number of filters in order to \cblue{be} compared with this reference. As an example, if an algorithm use $n_{ref}$ filters, \cblue{the} second part of \eqref{eq:IM} is equal to zero. Therefore, with $\text{NZ}=1$ (i.e. all coefficients are non-zero), a number of filters smaller than $n_{ref}$ obtains a positive $\text{IM}$ (good interpretability), while using more filters than $n_{ref}$ produces negative $\text{IM}$ (poor interpretability). In this paper, we use the number of classes to be classified ($m$) as $n_{ref}$ ($n_{ref}=m$). Note that $\text{IM}$ increases linearly as $\text{NZ}$ approaches one logarithmically. Figure 5 displays the $\text{IM}$ curve in terms of $n_f$ and $\text{NZ}$ with $n_{ref}=10$.

 \begin{figure}[!t]
\centering
\includegraphics[width=3in]{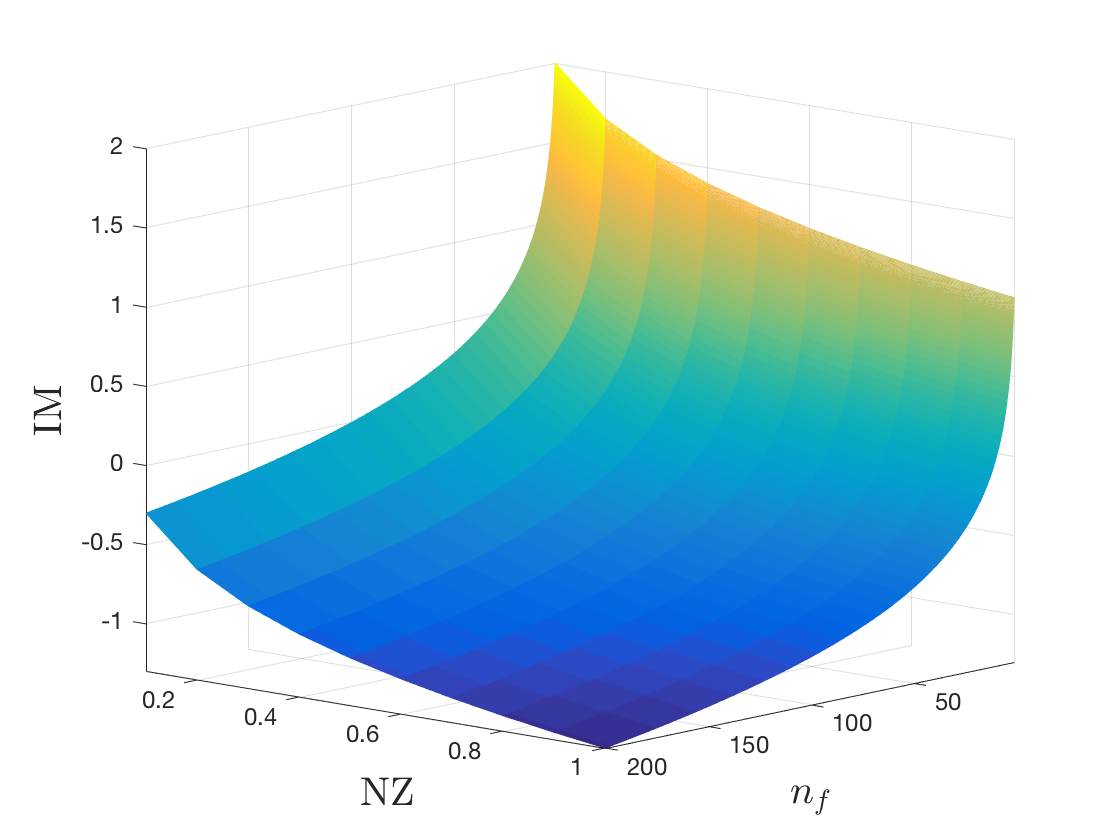}
\caption{IM curve in terms of $n_f$ and NZ.}
\label{Fig:interpretability}
\end{figure}

\subsection{Experiment 1: Texture Classification}

This subsection considers two different image texture classification tasks: classification based on a predefined set of categories, which is a more realistic scenario for texture classification, and \cblue{an image classification task}, which is also a typical task used in the literature.

\begin{figure*}[!t]
\centering
\includegraphics[width=5.5in]{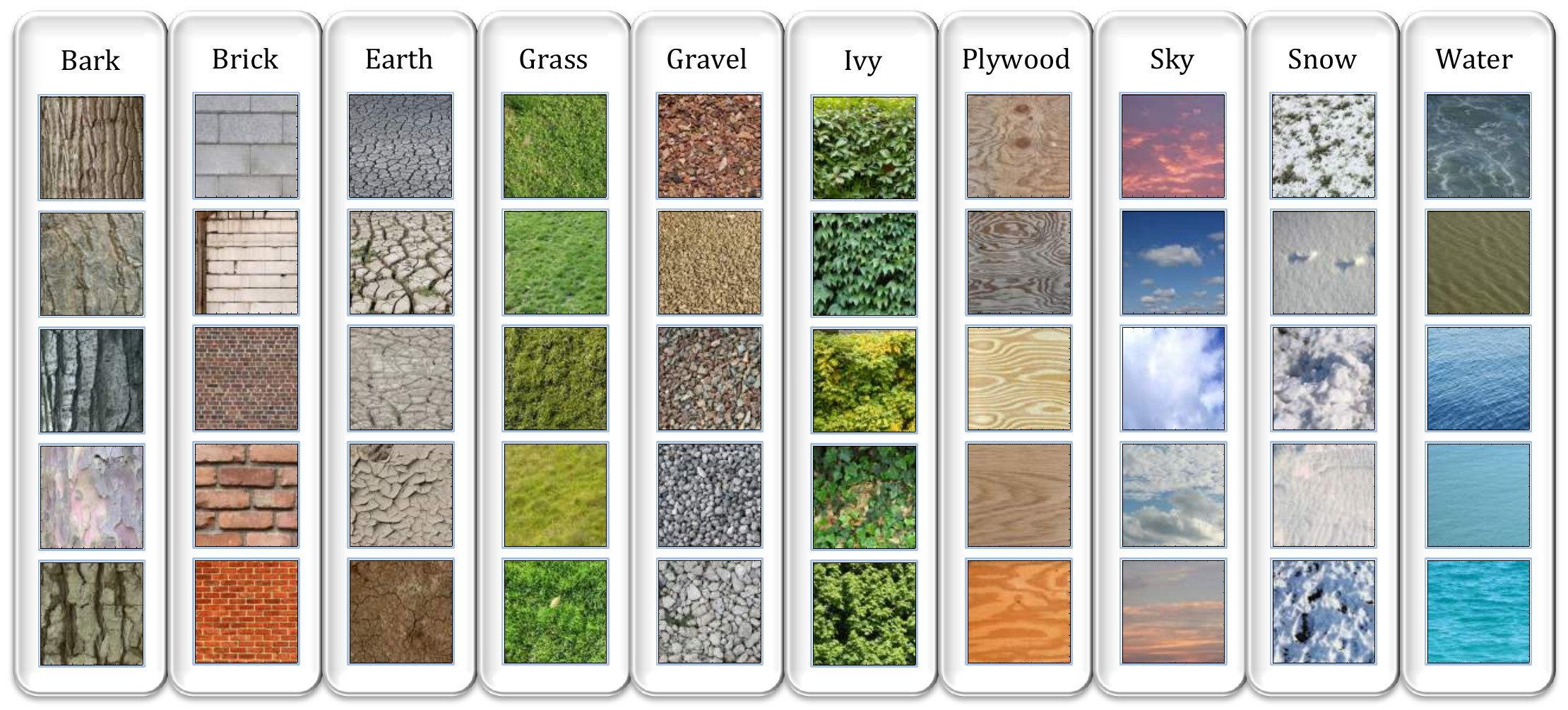}
\caption{A five images per class excerpt of the CGTextures dataset. In the preprocessing step, each of these images of size 480$\times$480 pixels is divided in 16 subimages of size 120$\times$120, which are the images used for the texture classification task.}
\label{Fig:database1}
\end{figure*}

The first task considers a real scenario for texture classification, where each image belongs to a specific class of textures\footnote{Textures were downloaded from \url{http://www.cgtextures.com/} in 2009 and the dataset created and used in this paper can be downloaded from \linebreak \url{http://www.tsc.uc3m.es/~smunoz/CGTextures.zip}. Due to the origin of the textures, we will refer to this dataset as CGTextures.} among 10 different categories: bark, earth, gravel, plywood, snow, brick, grass, ivy, sky, and water. In order to provide more patterns to the database, each image is divided in a set of 16 sub-images. The second task considers the  Brodatz dataset \cite{Brodatz1966} which has been widely used in the texture classification literature. In this experiment, each image is also divided in a set of 16 sub-images and the goal of the classification task consists  of assigning each sub-image to the original image. Table \ref{Tab:texture_datasets} summarizes the main characteristics of these datasets. Figure \ref{Fig:database1} displays a 5 images per class excerpt of the CGTextures dataset, where each class consists of very different pictures, making this a difficult texture classification task.

For the following experiments, we have divided each image of side-size $L=480$ pixels in 16 sub-images, and for our methods we have converted each sub-image into a frequency image of $12\times 12$ pixels (i.e., $\rho=12$) decimating the original frequency image by a factor of 10.

In case of \cblue{GF}, we also cross-validated (CV) the parameters $\eta$ and $\gamma$, setting their values to $\eta=0.5$ and $\gamma=0.5$ for both datasets. The  rest of parameters have been fixed according to \cite{Bianconi2007}: $n_F= 4$, $n_O=6$, and $F_r=\sqrt{2}$. Furthermore, the number of filters in the bank has been cross-validated for each method under study.

\begin{table}[!t]
\caption{Description of the main characteristics of the image data sets used for texture classification.}
\label{Tab:texture_datasets}
\centering
\begin{tabular}{@{}llll@{}}
\toprule
 & No. images (train/test) & Size & No. classes\\
\midrule
CGTextures & 3840/1568 & 120$\times$120 & 10\\
Brodatz \cite{Brodatz1966} & 1332/444 & 120$\times$120 & 111\\
\bottomrule
\end{tabular}
\end{table}

Thereupon, we will study the discriminative power and interpretability of the proposed supervised filter designs in comparison to the \cblue{specifically} designed \cblue{GF} banks \cite{Bianconi2007}. After designing each filter bank, we will train a linear Support Vector Machine (C-SVM) using the projected input data ($\XX'=\UU\XX$) in order to evaluate the overall accuracy (OA) of every method; the optimal value of \cblue{parameter $C$, which sets a trade off between the classifier margin and the number of errors,} has been cross-validated for each method under study. \cblue{Since the aim of this paper is to obtain a subset of interpretable features useful for these classification tasks, 
the interpretability capability will be analyzed by measuring the number of frequencies used by each filter bank and displaying the resulting projected data.}

\subsection*{Texture Classification in the CGTextures dataset}

\begin{figure*}[!t]
\centering
\begin{tabular}{cc}
\includegraphics[width=2.8in]{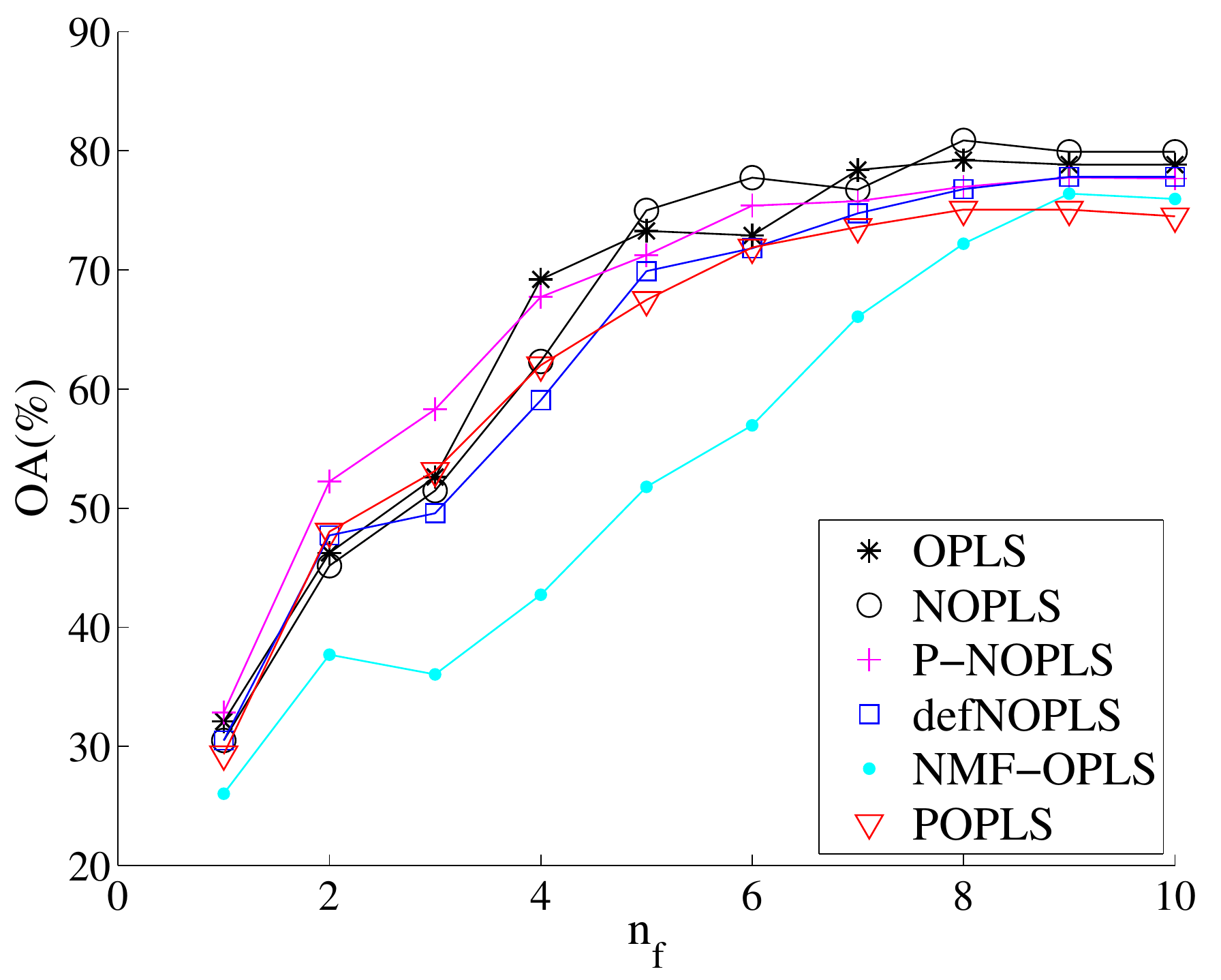} & \includegraphics[width=2.8in]{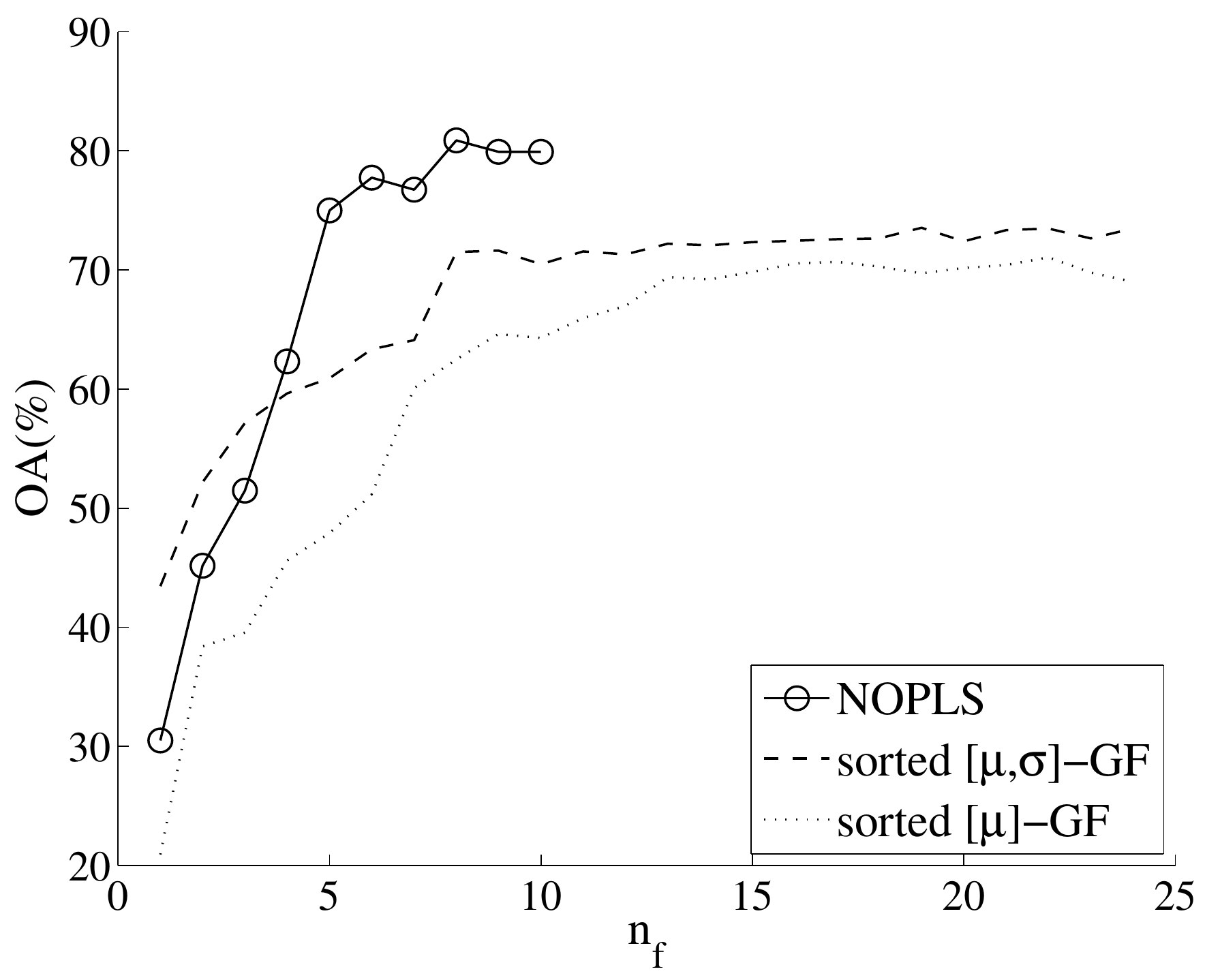}\\
a) & b)
\end{tabular}
\caption{Performance comparison \cblue{for the CGTextures dataset} among (a) our methods and (b) best-performing NOPLS method and \cblue{GF} bank using mean and standard deviation ($[\mu,\sigma]$) or the mean only ($[\mu]$) of each filtered image.}
\label{Fig:OA_database1}
\end{figure*}

Table \ref{Tab:database1_comparison} and  Figure \ref{Fig:OA_database1} compare the performances obtained by the proposed methods and by the \cblue{GF bank} over the CGTextures dataset. In particular, Figure \ref{Fig:OA_database1} displays the evolution of the overall accuracy (OA) with the number of filters in the bank, and Table \ref{Tab:database1_comparison} shows the OA of each method when $n_f$ is selected using CV. To carry out a fair analysis, we have included results for the GF approach sorting out the filters according to MSE in the training set, i.e., for each $n_f$ we selected the subset of filters achieving the best recognition capabilities.



\begin{table}[!t]
\caption{Performance comparison among our proposals and sorted \cblue{GF} for the CGTextures dataset.}
\label{Tab:database1_comparison}
\centering
\begin{tabular}{@{}llllll@{}}
\toprule
Algorithm & OA($\%$) & $n_f$ & $\#feat.$ & NZ & IM\\
\midrule
NOPLS & $\mathbf{79.91}$ & 9 & 9 & 0.046 &  1.4 \\
P-NOPLS & 77.74 & 10 & 10 & 0.029 & 1.5 \\
defNOPLS & 77.81 & 9 & 9 & 0.041 &  1.4 \\
NMF-OPLS & 75.96 & 10 & 10 & 0.045 &  1.3 \\
POPLS & 74.49 & 10 & 10 & 0.031 &  1.5 \\
OPLS & 79.21 & 8 & 8 & 0.800 &  0.2 \\
sorted GF & 73.47 & 24 & 48 & 0.524 &  0.4 \\
\bottomrule
\end{tabular}
\end{table}

As expected, the set of proposed supervised filter designs, and mainly the NOPLS algorithms, present an increased accuracy with respect to GF approaches; note that NOPLS outperforms the rest of the algorithms including OPLS. 
Moreover, the number of filters used by the supervised filter banks is less than half the number of filters selected for the GF bank. Furthermore, it is important to remark that, although all methods use a similar number of filters ($n_f$), the number of frequency bands selected by our methods is significantly smaller than by GF as it can be seen with the rate of non-zero coefficients of the filters (NZ) in Table \ref{Tab:database1_comparison}.

Besides, as we explained in Section \ref{section:textures}, our methods only extract one feature by each filter (see $\#feat.$ in Table \ref{Tab:database1_comparison}), while GF uses two features extracted by each filter: the mean of the filtered image ($\mu$), and its standard deviation ($\sigma$). In order to compare the performance between the GF with one or two features by filter and our best method, we display a comparison of the evolution of the OA with the number of considered filters in Figure \ref{Fig:OA_database1} (b).

\begin{figure*}[!t]
\centering
\includegraphics[scale=.75]{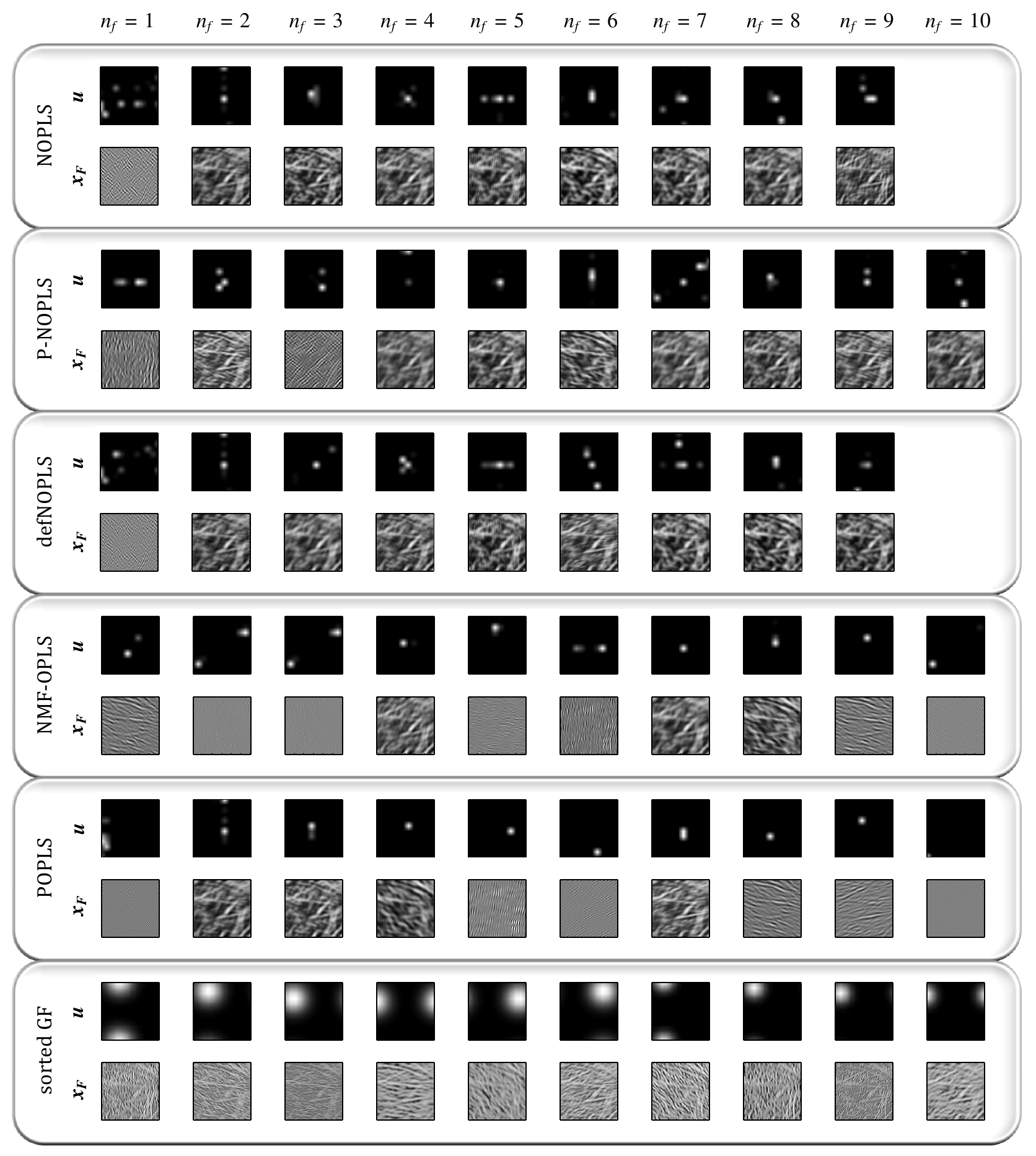}
\caption{Representation of the frequency response ($\uz$) of the 10 first filters used by each method in the texture classification task. \cblue{Note that NOPLS and defNOPLS use only 9 filters due to CV selection.} The corresponding filtered images ($\xz_F$) for an example of class {\em grass} have also been represented for the different methods and filters.\label{Fig:filteredImages}}
\end{figure*}

In summary, we can state that the proposed methods are more discriminative, more selective, and more sparse than GF. 
In order to analyze the interpretability of each method under study in a qualitative way, Figure \ref{Fig:filteredImages} shows the 10 first filters ($\uz$) of the filter bank provided by each method, as well as an example of the filtered images ($\xz_F=\xz\ast\uz$, being $\ast$ the convolution operation) from an image of the class \textit{grass}. As we can see, the supervised filters are \cblue{more accurate} and selective than those of the GF bank, being a mixture of band-pass filters in horizontal, vertical, and oblique directions. It is interesting to note the similarity among the filters in the banks of NOPLS, defNOPLS, and even the first filters of POPLS, which is indicative of NOPLS doing better than P-NOPLS. With respect to the GF bank, the worse accuracy of the classification system relying on these features \cblue{indicates} that this set of filters failed to extract discriminative features for the task at hand, \cblue{making clear the} convenience of designing the filter bank in a supervised manner.

\subsection*{Texture Classification in the Brodatz dataset}
In this subsection, we evaluate the different methods under study over the Brodatz texture classification scenario. In this case, each subimage has to be assigned to its original image correctly and, thus, the number of classes to be labeled is the same as the original number of images available in the database. As in the previous subsection, we compare our proposals with the GF bank, although in this case we use the GF design proposed in  \cite{Bianconi2007} where the GF bank is \cblue{specifically} designed for this particular task.

\begin{figure}[!t]
\centering
\includegraphics[width=3in]{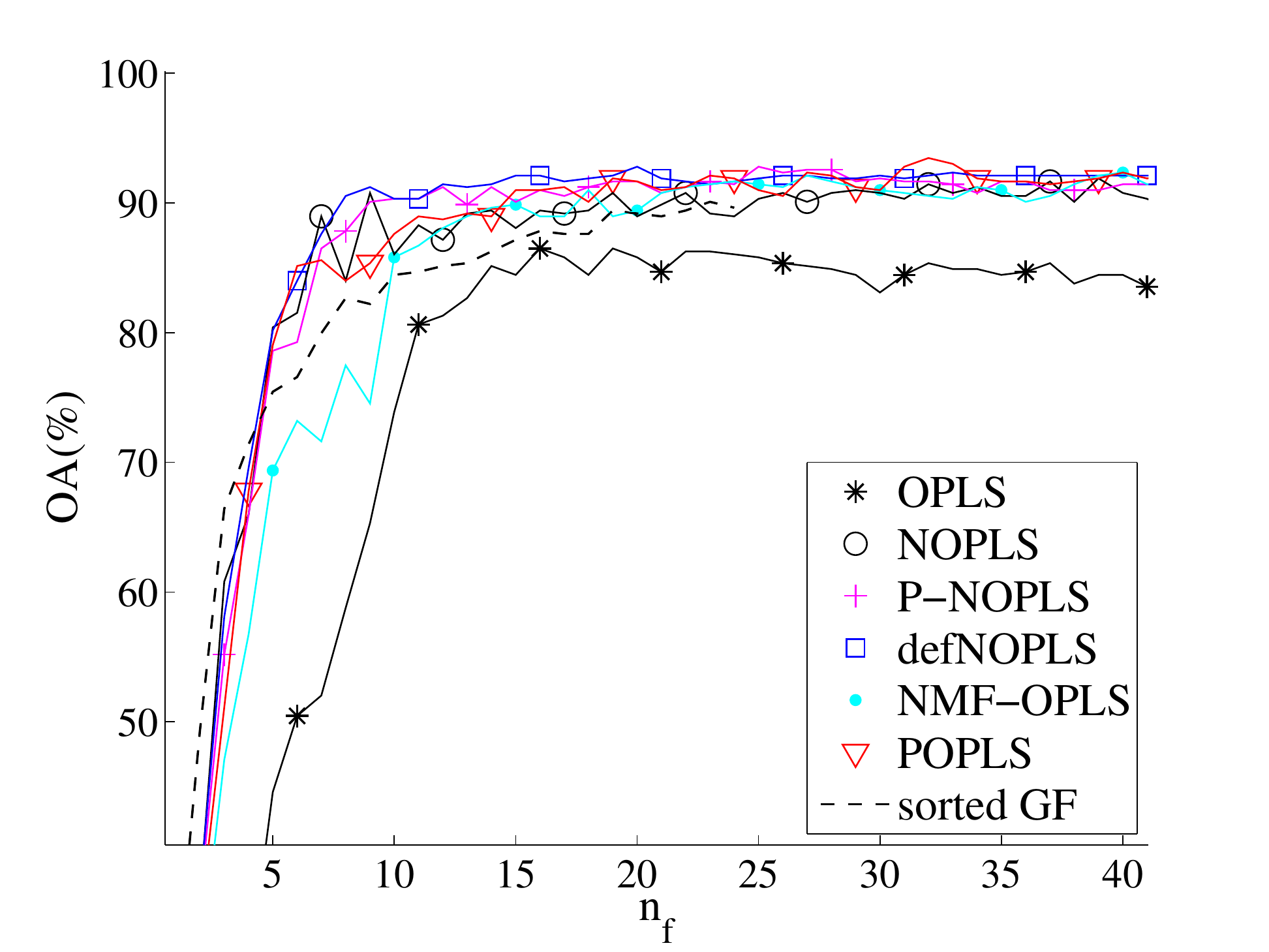}
\caption{Performance comparison among our proposals and GF for \cblue{the} Brodatz dataset. These curves display the OA as a function of the number of filters used in the filter bank ($n_f$).}
\label{Fig:OA_Brodatz}
\end{figure}

Following the same experimental procedure than in the previous experiment, Table \ref{Tab:Brodatz_comparison}  and Figure \ref{Fig:OA_Brodatz} include a comparison of the different methods under study. Again, to get a more fair comparison, the filters in the GF bank have been sorted according to \cblue{the MSE} measured over the training dataset. To further discuss the differences between the proposed methods, in this subsection we will analyze the training times required to build the filter banks.  All simulations were run using Matlab 8 on a Macbook Pro with 8 GB RAM Memory and a 2.9 GHz Dual-core Intel Core i7 CPU.

As we can see, all supervised methods are more discriminative than \cblue{the GF bank}, even when there are few filters in the banks. Although P-NOPLS is slightly more discriminative than NOPLS, it takes much longer to train, and the number of filters is also larger. It is important to remark that NOPLS is the fastest algorithm (2.34 sec.) --excluding the OPLS algorithm since it does not include any constraint on the formulation-- and requires half of the features than GF, whereas, in this case, the defNOPLS solution is the most discriminative and second fastest (14.84 sec.). P-NOPLS and NMF-OPLS need around 20 sec. and POPLS is dramatically slower with 12 h. and 12 min. \cblue{Unlike the previous problem}, GF here uses \cblue{less filters} than supervised approaches, however the number of coefficients \cblue{is} similar (except for P-NOPLS) and the number of frequency bands of the images needed by our algorithms are drastically smaller than GF (see NZ in Table \ref{Tab:Brodatz_comparison}). Comparing to OPLS results, we can see that the standard OPLS solution obtains the worst performance with any subset of filters. This fact points out as the non-negativity constraints not only provides interpretable solutions but also (in some cases) increased performance.

Note that, as we explained in Section \ref{section:designFilterBanks}, P-NOPLS and NMF-OPLS do not sort the filters of the bank in terms of their importance. One of the consequences of this is that they require more filters than the other supervised methods; as an example, we see that P-NOPLS needs twice the number of filters than the rest of \cblue{the} methods.

\begin{table}[!t]
\caption{Performance comparison among our proposals and sorted GFs \cblue{for the Brodatz dataset}.}
\label{Tab:Brodatz_comparison}
\centering
\begin{tabular}{@{}lllllllll@{}}
\toprule
Algorithm & OA($\%$) & $n_f$ & $\#feat.$ & NZ & IM & Time (sec.)\\
\midrule
NOPLS & 90.32 & 24 & 24 & 0.015 & 2.5 & \cblue{2.34}\\
P-NOPLS & 91.22 & 105 & 105 & 0.016 & 1.8 & \cblue{20.43}\\
defNOPLS & $\mathbf{92.12}$ & 53 & 53 & 0.0182 & 2.1 & \cblue{14.84}\\
NMF-OPLS & 90.99 & 63 & 63 & 0.009 & 2.3 & \cblue{19.87}\\
POPLS & 91.67 & 55 & 55 & 0.0059 & 2.5 & \cblue{43,948}\\
OPLS & 85.81 & 20 & 20 & 0.1802 & 1.5 & \cblue{0.01}\\
sorted GF & 90.09 & 23 & 46 & 0.5202 & 1.0 & -\\
\bottomrule
\end{tabular}
\end{table}

\subsection{Experiment 2: Music Genre Classification}
\label{subsection:genreClassification}

This second block of experiments aims at classifying \cblue{the} music genre of a song from the periodogram of the 6 first \cblue{MFCCs} extracted from each song. The dataset used here has been previously investigated in \cite{ArenasPOPLS,Meng2007,Meng2005}, and their results have revealed a great difficulty to successfully classify each song according to its musical genre (see \cite{ArenasPOPLS,Meng2005}). Moreover, the human evaluation study of \cite{Meng2005} has found that the human definition of the genre for the audios in this dataset presents low consistency, resulting in a difficult dataset to apply \cblue{genre classification}. Notwithstanding the above, it is interesting to study how supervised filter bank designs work in this setup.

The dataset consists of 1317 music snippets of 30s each, distributed evenly among the following 11 music genres: Alternative, Country, Easy Listening, Electronica, Jazz, Latin, Pop$\&$Dance, Rap$\&$Hiphop, R$\&$B and Soul, Reggae\cblue{,} and Rock. In \cblue{the} case of \cblue{the} Latin category, there are only 117 music samples. The music snippets are MP3 encoded music with a bitrate of 128 kbps or higher downsampled \cblue{by a factor of two} to 22050 Hz. Note that this dataset has on average 1.83 songs per artist, which is \cblue{another reason for} making it so hard for genre classification.

For comparison purposes, we are going to consider the Philips Filter Bank proposed in \cite{Mckinney2003} for a music genre classification task. As we explained at the end of Section IV, we will use periodograms of length D=129, so that the size of \cblue{ matrix $\UU$} characterizing the Philips \cblue{Filter Bank}, as well as the supervised banks designed with any of our methods, will be $129 \times n_f$, with $n_f = 4$ for the Philips \cblue{Filter Bank}. 

Due to the lack of a specific test subset, we apply a 10-fold cross-validation procedure in order to measure the classification accuracy of every method. In each fold, we design the optimal filters with nine partitions of the data, as described in Section \ref{section:music_genre}, and subsequently we evaluate the performance on the remaining partition. Note that all samples of the same song cannot be divided into different partitions, i.e., partitions are defined in terms of songs instead of samples of the dataset.

\begin{figure*}[!t]
\centering
\begin{tabular}{cc}
\includegraphics[width=2.5in]{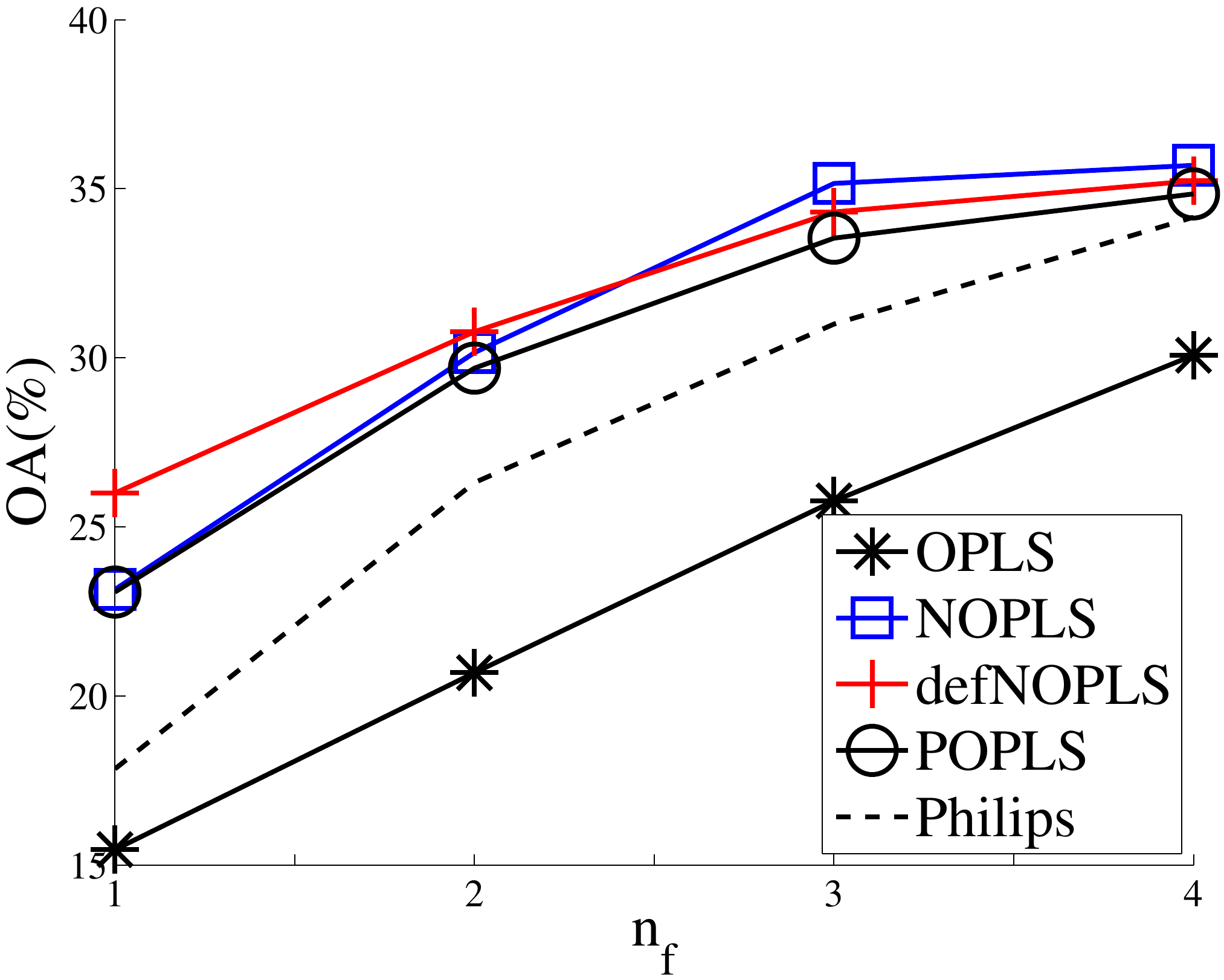} & \includegraphics[width=2.5in]{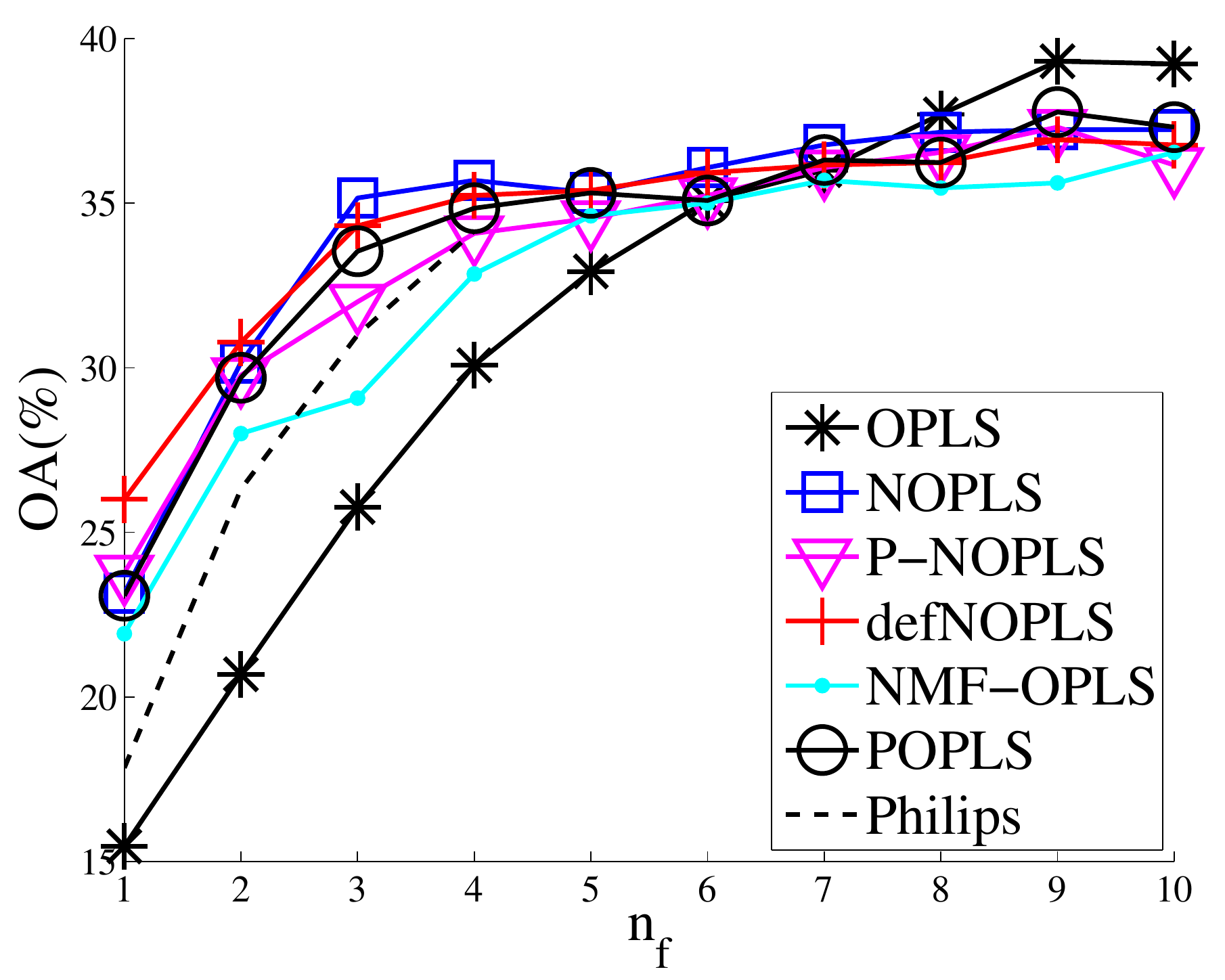}\\
a) & b)
\end{tabular}
\caption{Overall Accuracy (OA) comparison with (a) a detailed study among the best supervised filter banks and the Philips \cblue{Filter Bank} (only first 4 filters), and (b) a complete comparison among all methods with the full filter bank.}
\label{Fig:OA_genre}
\end{figure*}

A comparison among the supervised filter approaches and the Philips Filter Bank is displayed in Table \ref{Tab:OA_genre}. This table shows the OA (averaged over the 10 folds) when the 4 and 10 first filters in the bank are considered ($n_f=4$ and $n_f=10$, respectively). In the case of the Philips \cblue{Filter Bank}, results are only analyzed with 4 filters, since this is its maximum number of available filters. Table \ref{Tab:OA_genre} also includes the rate of non-zero coefficients of the filters (NZ) as well as the time required to design the different filter banks. To complete this analysis, Figure \ref{Fig:OA_genre} displays the averaged OA as a function of the number of filters for all methods under study. 

As we explained in Section \ref{section:designFilterBanks}, two of our proposed methods (P-NOPLS and NMF-OPLS) lack the ability to sort the filters in the bank with respect to the importance of each filter. As a consequence of this shortcoming, when only a few filters are used the performance may be adversely affected, as is the case here, where these methods are even outperformed by the Philips \cblue{Filter Bank} when $n_f=4$. Regarding the remaining supervised filters, it is not so clear which one presents the best performance: although POPLS has the best accuracy with $n_f=10$, NOPLS obtains a similar performance, but with a lower percentage of non-zero coefficients. Even more, the accuracy of both defNOPLS and NOPLS are the best ones when few filters are used (see left subfigure of Figure \ref{Fig:OA_genre}), improving significantly the performance of the Philips \cblue{Filter Bank}. One of the advantages of defNOPLS with respect to NOPLS, is that the sequential implementation of the algorithm allows to automatically select $n_f$ and stop the training process.

\begin{figure*}[!t]
\centering
\includegraphics[scale=0.7]{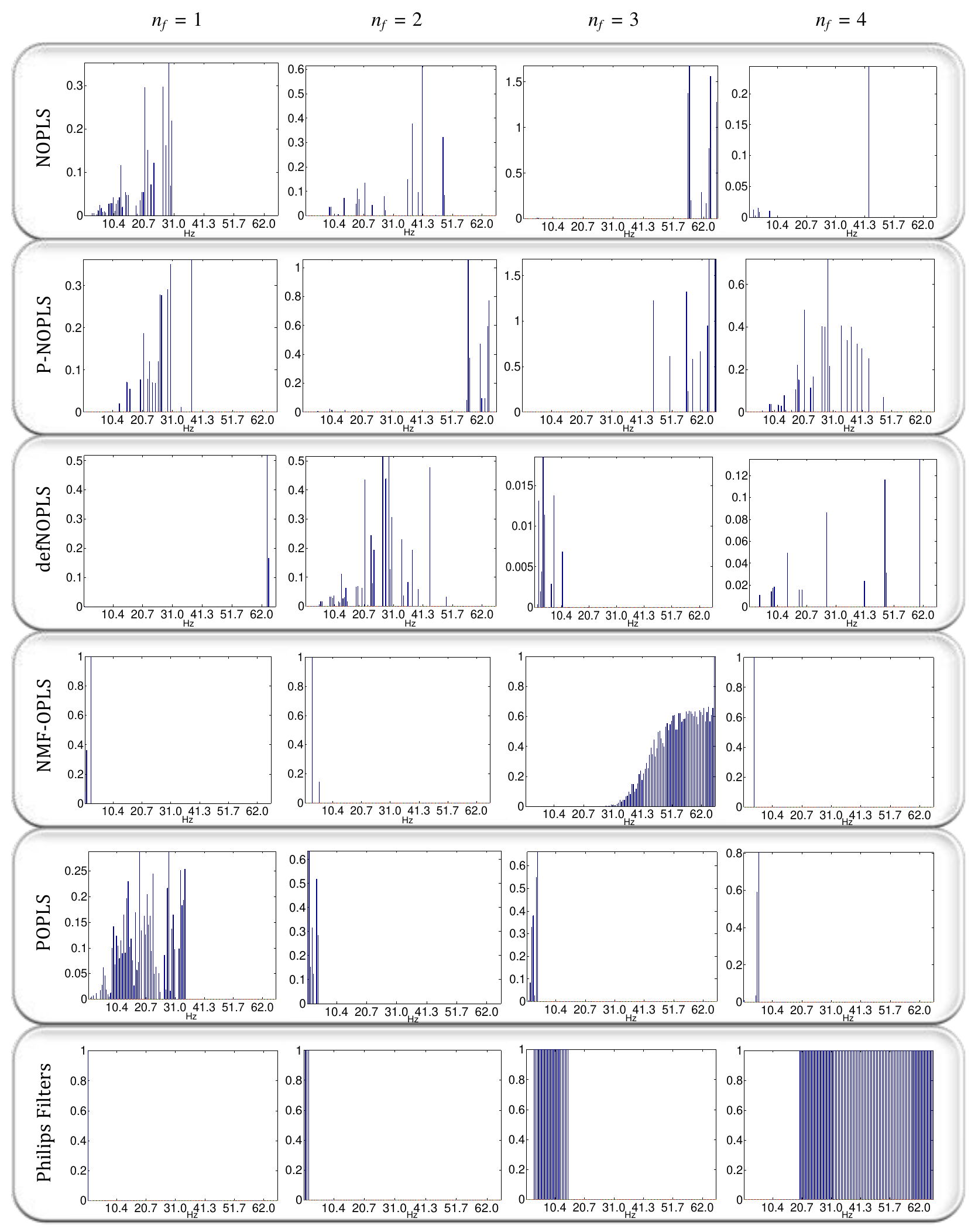}
\caption{Frequency response of the four first filters designed by each algorithm.\label{Fig:frecuencyFilters}}
\end{figure*}

To conclude the section, Figure \ref{Fig:frecuencyFilters} shows the first 4 filters obtained on a single fold\footnote{We have checked that the differences between the filters obtained for each data fold are not very significant, so the presented conclusions can be easily generalized to the remaining folds.} for the first MFCC, so that we can analyze the information provided by each filter bank. Note that, similarly to the Philips \cblue{Filter Bank}, NOPLS, defNOPLS, and POPLS pay attention to three well differentiated regions of the spectra, even though they are not in the same order: the lower modulation frequencies, which \cblue{include} components at the beat-scale; the higher modulation frequencies, which are related to the perceptual roughness; and the modulation frequencies of instruments, which are the most important frequencies of the MFCCs periodograms.
 However, the supervised approaches are more flexible in the definition of the filters, and can adjust the cut-off frequencies and even shape the filter waveform to obtain the best possible performance in the genre classification task. The superior performance of the supervised techniques allows us to conclude the convenience of using the available target data not only for the training of the final classifier, but also in the design of the filters used in the feature extraction stage.


\begin{table}[!t]
\caption{OA ($\%$) in the genre classification task when using $n_f = 4$ and $n_f = 10$ filters. The percentage of non-zero coefficients and training time required by each method are also displayed.}
\label{Tab:OA_genre}
\centering
\begin{tabular}{@{}lcclcll@{}}
\toprule
Algorithm & \hspace{-.4cm}\begin{tabular}{c}OA\\($n_f=4$)\end{tabular} & \hspace{-.6cm}\begin{tabular}{c}OA\\($n_f=10$)\end{tabular} & \hspace{-.6cm}\begin{tabular}{c}NZ\\\cblue{($n_f=4$)}\end{tabular} & \hspace{-.6cm}\begin{tabular}{c}IM\\($n_f=4$)\end{tabular} & \hspace{-.2cm}Time (sec.)\\
\midrule
NOPLS & \hspace{-.4cm}\textbf{35.69} & \hspace{-.6cm}37.23 & \hspace{-.2cm}\textbf{0.029} &  \hspace{-.6cm} 2.0    & \hspace{-.2cm}6.56\\     
P-NOPLS & \hspace{-.4cm}34.07 & \hspace{-.6cm}36.15 & \hspace{-.2cm}0.167 &  \hspace{-.6cm}    1.2    & \hspace{-.2cm}7.65\\     
defNOPLS & \hspace{-.4cm}35.23 & \hspace{-.6cm}36.77 & \hspace{-.2cm}0.039 &  \hspace{-.6cm}    1.8   & \hspace{-.2cm}15.40\\     
NMF-OPLS & \hspace{-.4cm}32.85 & \hspace{-.6cm}36.54 & \hspace{-.2cm}0.063 &   \hspace{-.6cm}    1.6   & \hspace{-.2cm}32.13\\     
POPLS & \hspace{-.4cm}34.85 & \hspace{-.6cm}37.31 & \hspace{-.2cm}0.138 &   \hspace{-.6cm}    1.3   & \hspace{-.2cm}2667.59\\     
OPLS & \hspace{-.4cm}30.08 & \hspace{-.6cm}\textbf{39.23} & \hspace{-.2cm}1.000 &  \hspace{-.6cm}    0.4   & \hspace{-.2cm}\textbf{2.7}\\
Philips F. B. & \hspace{-.4cm}34.15 & \hspace{-.6cm}- & \hspace{-.2cm}0.038 &   \hspace{-.6cm}    1.9   & \hspace{-.2cm}-\\
\bottomrule
\end{tabular}
\end{table}

\subsection{Experiment 3: Comparison with deep learning solutions}
\label{subsection:deepLearningComparison}

To complete this experimental analysis, in this section we compare the performance of the proposed methods with deep-learning approaches, since they are considered the \cblue{state-of-the-art} in image recognition. With this purpose, we have selected these Deep Neural Networks (DNN):
\begin{itemize}
\item T-CNN-3 [59], which is a 3-layer convolutional neural \cblue{network} (CNN), specifically designed for texture recognition problems,
\item AlexNet [60], which is probably one the most popular \cblue{state-of-the-art} DNN architectures,
\item FV-CNN [61], that combines a CNN architecture with an internal representation based on the so-called Fisher vectors, and is the current \cblue{state-of-the-art} solution in texture classification,
\end{itemize}
and we will compare their performance with our proposed methods in two texture classification databases: kth-tips-2b \footnote{\url{https://github.com/v -andrearczyk/caffe-TCNN}} [62] and DTD \footnote{\url{https://www.robots.ox.ac.uk/~vgg/data/dtd/}} [63], with 11 and 47 classes, respectively. In order to fairly compare all the algorithms, the same training and test data partitions have been used for all the methods and as well as the same experimental configuration as [59].

The results for all DNN methods have been directly taken from [59]. As we can see in Table \ref {Tab:OA_DNN}, our methods \cblue{outperform} these \cblue{state-of-the-art} techniques. Regarding the interpretability of the different solutions, we have reported the objective measure IM for our approaches. In spite of their good performance, the main criticism of DNNs is precisely that they can be seen as black boxes implementing a very large number of connections among computation nodes. Consequently, the number of filters is very large, which results in small IM values. To be more precise, and using the information available in [60], AlexNet is using 613,120 \cblue{filters} for a final IM of \cblue{-4.7 and -4.1 in kth-tips-2b and DTD, respectively}. For \cblue{the} T-CNN-3 method, the number of units reported in [59] is 1,824, and thus it obtains a final IM of \cblue{-2.2 and -1.6 in kth-tips-2b and DTD, respectively}. IM of these DNN approaches is significantly smaller than for our methods. In summary, our methods show competitive performance in these problems, while providing more interpretable solutions.

For completeness, we should also mention that pretrained DNNs have also been used in [59,61], where the widely-used ImageNet database has been used as a pre-step to adjust the intermediate units of the networks. \cblue{Although this approach is very useful from a practical point of view,} these results cannot be directly compared to our methods, since for a fair comparison we should design specific procedures that allow our feature extractors to learn from additional databases such as ImageNet, which is out of the scope of this paper. Nevertheless, our methods still remain better than the results provided for DNN approaches in \cblue{the} kth-tips-2b dataset, being 73.2\%, 71.5\% and 81.5\% for T-CNN-3, AlexNet and FV-CNN, respectively. However, using ImageNet together with the DTD problem allows deep learning networks to significantly improve their performance, achieving up to 75\% for the current \cblue{state-of-the-art} FV-CNN method, which is much better than any of the results in Table \ref{Tab:OA_DNN} for that database. \cblue{This suggests that, independently of the selected method, certain databases (such as DTD) probably do not contain enough diversity to make possible the design of a sufficiently good set of filters, and using external databases can be rather beneficial.} 

\begin{table}[!t]
\caption{OA ($\%$) and IM comparison with \cblue{state-of-the-art} deep learning approaches in  texture classification tasks. \cblue{FV-CNN and pretrained DNN approaches using Imagenet results are reported in the text.}}
\label{Tab:OA_DNN}
\centering
\begin{tabular}{@{}lcccc@{}}
\toprule
Algorithm &  \multicolumn{2}{c}{kth-tips-2b} & \multicolumn{2}{c}{DTD}\\
 & OA & IM & OA & IM\\
\midrule
NOPLS & 88.4 & 1.3 & 29.1 & 1.4 \\
P-NOPLS &  84.3 & 1.3 & \textbf{31.8} & 1.7 \\
defNOPLS & \textbf{88.6} & 1.3 & 31.1 & 1.5 \\
NMF-OPLS & 71.6 & 1.1 & 28.6 & 1.1 \\
POPLS & 79.1 & 1.5 & 31.3 & 1.9 \\
T-CNN-3 & 48.7 & -2.2 & 27.8 & -1.6\\
AlexNet & 47.6 & -4.7 & 22.7 & -4.1\\
\bottomrule
\end{tabular}
\end{table}

\begin{figure*}[t]
\centering
\includegraphics[scale=.18]{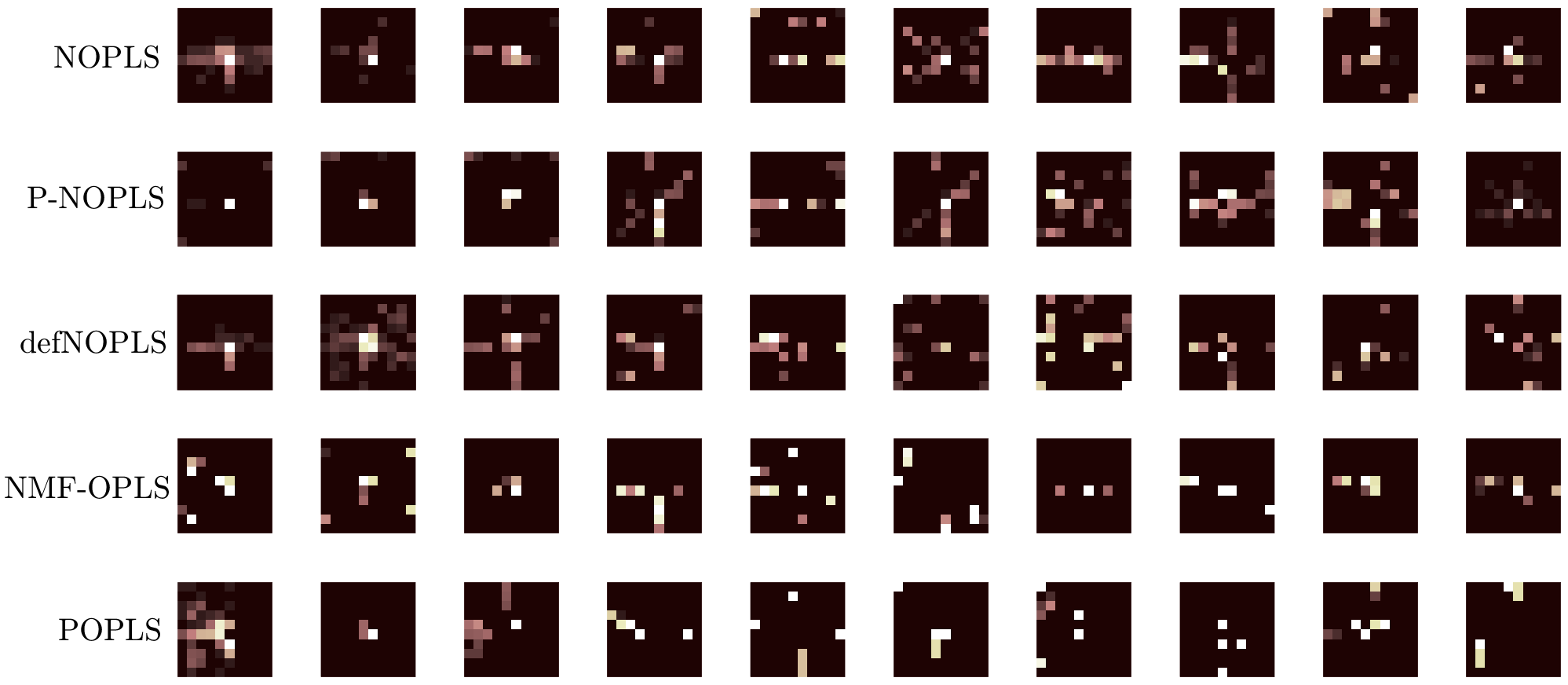}
\caption{Filter bank obtained as an average of the 4 filter banks from the 4 given splits by using NOPLS in \cblue{the} kth-tips-2b dataset. \cblue{Note that $n_f=10$ has been selected for all of these methods by using CV prodecure.}}
\label{Fig:filterBank_kthtips2b}
\end{figure*}

Regarding interpretability, Figure \ref{Fig:filterBank_kthtips2b} shows that just 10 sparse filters can be very useful for image processing experts to understand the filter design. In the case of DNN approaches, we have not considered showing their filters due to the high number of obtained filters, which are also not ordered by any criterion of relevance to the classification.

\section{Conclusions}
\label{section:conclusions}
In this paper, we have presented different methods for designing very versatile and interpretable filter banks for particular visual or audio classification tasks. All proposed methods are based on a supervised design with a common objective function, and differ in the way they try to solve this non-convex problem. As an alternative to the POPLS algorithm proposed in \cite{ArenasPOPLS}, in this paper we propose several far less time-consuming methods which obtain similar or even better performance than POPLS. Moreover, our proposals outperform the \cblue{purposely designed} and well studied filter banks used in the state-of-art of visual and audio applications.

We have illustrated the versatility of our methods in our experiments section, where we have tackled two very different classification tasks: texture and music genre classification. The advantages of our approaches over other feature extraction methods are: 1) they provide elegant physical interpretations of the extracted features; 2) they are more discriminative with less number of filters; 3) they provide more interpretable and sparse solutions; 4) and they fit their filter banks to each particular task, unlike generic filter banks. Based on our findings, we can conclude that the block and deflated NOPLS algorithms seem to obtain the best results in terms of accuracy, sparsity and CPU requirements and, therefore, they should \cblue{be} preferred over the other methods.

%
%

\ifCLASSOPTIONcaptionsoff
  \newpage
\fi



%

\bibliographystyle{IEEEtran}
\end{document}